\documentclass[twocolumn,showpacs,preprintnumbers,amsmath,amssymb,floatfix]{revtex4}

\usepackage{verbatim}
\usepackage{graphicx}
\usepackage{dcolumn}
\usepackage{bm}
\usepackage{color}
\usepackage{url}
\usepackage{amsmath}
\usepackage{morefloats}

\usepackage{color}

\newcommand{\RS}[1]{{\bf\color{blue}{#1}}}

\graphicspath{{./figures/}}
\newcommand{\be}{\begin{equation}}
\newcommand{\ee}{\end{equation}}
\newcommand{\ba}{\begin{eqnarray}}
\newcommand{\ea}{\end{eqnarray}}
\newcommand{\nn}{\nonumber}
\newcommand{\Mc}{{\cal M}}

\newcommand{\hyp}{\mathcal{H}}
\newcommand{\info}{\mathrm{I}}
\def\ltsima{$\; \buildrel < \over \sim \;$}
\def\simlt{\lower.5ex\hbox{\ltsima}}
\def\gtsima{$\; \buildrel > \over \sim \;$}
\def\simgt{\lower.5ex\hbox{\gtsima}}
\begin{document}

\title[blah]{Towards a generic test of the strong field dynamics of general relativity using compact binary coalescence}

\author{T.G.F.~Li$^{1}$, W.~Del Pozzo$^{1}$, S.~Vitale$^{1}$, C.~Van Den Broeck$^{1}$, M.~Agathos$^{1}$, J.~Veitch$^{1,2}$, K.~Grover$^{3}$, T.~Sidery$^{3}$, R.~Sturani$^{4,5}$, A.~Vecchio$^{3}$}
\affiliation{$^1$Nikhef -- National Institute for Subatomic Physics, Science Park 105, 1098 XG Amsterdam, The Netherlands \\
$^2$School of Physics and Astronomy, Cardiff University, Queen's Buildings, The Parade, Cardiff CF24 3AA, United Kingdom\\
$^3$School of Physics and Astronomy, University of Birmingham, Edgbaston, Birmingham B15 2TT, United Kingdom\\
$^4$Dipartimento di Scienze di Base e Fondamenti, Universit\`a di Urbino, I-61029 Urbino, Italy\\
$^5$INFN, Sezione di Firenze, I-50019 Sesto Fiorentino, Italy}
\date{\today}

\begin{abstract}
Coalescences of binary neutron stars and/or black holes are amongst the most likely gravitational-wave signals to be observed in ground based interferometric detectors. Apart from the astrophysical importance of their detection, they will also provide us with our very first empirical access to the genuinely strong-field dynamics of General Relativity (GR). We present a new framework based on Bayesian model selection aimed at detecting deviations from GR, subject to the constraints of the Advanced Virgo and LIGO detectors. The method tests the consistency of coefficients appearing in the waveform with the predictions made by GR, without relying on any specific alternative theory of gravity. The framework is suitable for low signal-to-noise ratio events through the construction of multiple subtests, most of which involve only a limited number of coefficients. It also naturally allows for the combination of information from multiple sources to increase one's confidence in GR or a violation thereof. We expect it to be capable of finding a wide range of possible deviations from GR, including ones which in principle cannot be accommodated by the model waveforms, on condition that the induced change in phase at frequencies where the detectors are the most sensitive is comparable to the effect of a few percent change in one or more of the low-order post-Newtonian phase coefficients. In principle the framework can be used with any GR waveform approximant, with arbitrary parameterized deformations, to serve as model waveforms. In order to illustrate the workings of the method, we perform a range of numerical experiments in which simulated gravitational waves modeled in the restricted post-Newtonian, stationary phase approximation are added to Gaussian and stationary noise that follows the expected Advanced LIGO/Virgo noise curves.
\end{abstract}


\maketitle

\section{Introduction}

General Relativity (GR) is a non-linear, dynamical theory of gravity. Until the 1970s, all of its tests involved the weak-field, stationary regime; these are the standard Solar System tests that are discussed in most textbooks, \emph{e.g.}~\cite{MTW}. GR passed them with impressive accuracy. Nevertheless, the more interesting part of any field theory resides in its dynamics, and this is especially true of GR \cite{Will:2006,SathyaprakashSchutz:2009}. A first test of the latter came from the Hulse-Taylor binary and a handful of similar tight neutron star binaries \cite{ht75,b03,k06}, whose orbital elements are changing in close agreement with GR under the assumption that energy and angular momentum are carried away by gravitational waves (GW). Thus, these discoveries led to the very first, albeit indirect, evidence for GW. However, even the most relativistic of these binaries, PSR J0737-3039~\cite{b03,k06}, is still in the relatively slowly varying, weak-field regime from a GR point of view, with a compactness of $G M/(c^2 R) \simeq 4.4 \times 10^{-6}$, with $M$ the total mass, $R$ the orbital separation, and a typical orbital speed $v/c \simeq 2 \times 10^{-3}$. By contrast, for an inspiraling compact binary, in the limit of a test particle around a non-spinning black hole, the last stable orbit occurs at a separation of $R = 6 G M/c^2$, where $G M/(c^2 R) = 1/6$ and $v/c = 1/\sqrt{6}$. This constitutes the genuinely strong-field, dynamical regime of General Relativity, which in the foreseeable future will only be empirically accessible by means of gravitational-wave detectors.

Several large gravitational wave observatories have been operational for some years now: the two 4 km arm length LIGO interferometers in the US \cite{LIGO}, the 3 km arm length Virgo in Italy \cite{Virgo,virgo}, and the 600 m arm length GEO600 \cite{GEO600}. By around 2015, LIGO and Virgo will have been upgraded to their so-called advanced configurations \cite{aLIGO,advligo,aVirgo,advvirgo}, and shortly afterwards up to tens of detections per year are expected \cite{ratespaper}. Another planned GW observatory is the Japanese LCGT \cite{LCGT}, and the construction of a further large interferometer in India is under consideration \cite{IndiGO}. Among the most promising sources are the inspiral and merger of compact binaries composed of two neutron stars (BNS), a neutron star and a black hole (NSBH), or two black holes (BBH).

Within GR, especially the inspiral part of the coalescence process has been modeled in great detail using the post-Newtonian (PN) formalism (see \cite{PN} and references therein), in which quantities such as the conserved energy and flux are found as  expansions in $v/c$, where $v$ is a characteristic speed. During inspiral, the GW signals will carry a detailed imprint of the orbital motion. Indeed, the contribution at leading order in amplitude has a phase that is simply $2\Phi(t)$, with $\Phi(t)$ being the orbital phase. Thus, the angular motion of the binary is directly encoded in the waveform's phase, and assuming quasi-circular inspiral, the radial motion follows from the instantaneous angular frequency $\omega(t) = \dot{\Phi}(t)$ through the relativistic version of Kepler's Third Law. If there are deviations from GR, the different emission mechanism and/or differences in the orbital motion will be encoded in the phase of the signal waveform, allowing us to probe the strong-field dynamics of gravity. In this regard we note that with binary pulsars, one can only constrain the \emph{conservative} sector of the dynamics to 1PN order (\emph{i.e.}~$(v/c)^2$ beyond leading order), and the dissipative sector to leading order; see, \emph{e.g.}, the discussion in \cite{Maggiore} and references therein. Hence, when it comes to $\Phi(t)$, these observations do not fully constrain the 1PN contribution. Yet several of the more interesting dynamical effects occur starting from 1.5PN order; this includes `tail effects' \cite{bs94,bs95} and spin-orbit interactions; spin-spin effects first appear at 2PN \cite{bdiww95}. As indicated by the Fisher matrix results of \cite{mais10}, Advanced LIGO/Virgo should be able to put significant constraints on the 1.5PN contribution to the phase, and possibly also higher-order contributions.

Possible deviations from GR that have been considered in the past in the context of compact binary coalescence include scalar-tensor theories \cite{w94,de98,sw01,wy04,bbw04,bbw05}; a varying Newton constant \cite{yps10}; modified dispersion relation theories, usually referred to in literature as
`massive graviton' models \footnote{The designations `massive gravity' and `massive graviton' originate
from \cite{bdiww95}
 where actually the effect of a modified
dispersion relation, or a wavelength dependent propagation
velocity has been taken into account.
While it is attractive to ascribe such a modification to a graviton mass,
a modification of the dispersion relation can be a more general effect, and
moreover endowing the graviton with a mass introduces additional deviations
from GR than a mere modified dispersion relation. See \emph{e.g.} the original
\cite{vdvz} and the recent \cite{deRham:2010kj} for a thorough discussion of the
issues related to massive gravity models.}
\cite{w98,bbw04,bbw05,aw09,sw09,ka10,dpvv11,BertiGairSesana:2011}; violations of the No Hair Theorem \cite{bcw06,Hughes:2006,bc07,glm08,khhs11}; violations of Cosmic Censorship \cite{vdbs06,khhs11}; and parity violating theories \cite{afy08,yf09,sy09,yooa10}. The (rather few) \emph{specific} alternative theories of gravity that have been considered in the context of ground-based gravitational wave detectors -- essentially scalar-tensor and `massive graviton' theories -- happen to be hard to constrain much further with GW observations, and we will not consider them in this paper. However, General Relativity may be violated in some other manner, including a way that is yet to be envisaged. This makes it imperative to develop methods that can search for \emph{generic} deviations from GR.

In the past several years, several proposals have been put forward to test GR using coalescing compact binary coalescence:
\begin{enumerate}
\item One can search directly for the imprint of specific alternative theories, such as the so-called `massive graviton' models and scalar-tensor theories \cite{w94,de98,sw01,wy04,bbw04,bbw05,w98,aw09,sw09,ka10,BertiGairSesana:2011}. For the `massive graviton' case, a full Bayesian analysis was recently performed by Del Pozzo, Veitch, and Vecchio \cite{dpvv11}.
\item A method due to Arun \emph{et al.} exploits the fact that, at least for binaries where neither component has spin, all coefficients $\psi_i$ in the PN expansion (see Eq.~(\ref{psiidef}) below for their definition) of the inspiral phase depend only on the two component masses, $m_1$ and $m_2$ \cite{aiqs06a,aiqs06b,mais10}. In that case only two of the $\psi_i$ are independent, so that a comparison of any three of them allows for a test of GR. Such a method would be very general, in that one does not have to look for any particular way in which gravity might deviate from GR; instead it allows generic failures of GR to be searched for. However, so far its viability was only explored using Fisher matrix calculations.
\item In the so-called parameterized post-Einsteinian (ppE) formalism of Yunes and Pretorius, gravitational waveforms are parameterized so as to include effects from a variety of alternative theories of gravity \cite{yp09,yh10}. A Bayesian analysis was performed in \cite{csyp11}.
\item Recently a test of the No Hair Theorem was presented, also in a Bayesian setting \cite{gvs11}.
\end{enumerate}

The first method presupposes that GR will be violated in a particular way. As for the third and fourth points, the particular Bayesian implementation that was used involves comparing the GR waveform with a waveform model that includes parameterized deformations away from GR, thus introducing further free parameters. 

Now, imagine one introduces free parameters $p_i$, $i = 1, 2, \ldots, N_T$ in such a way that $p_i = 0$ for all $i$ corresponds to GR being correct. Then one can compare a waveform model in which all the $p_i$ are allowed to vary with the GR waveform model in which all the $p_i$ are zero. As we will explain in this paper, this amounts to asking the question: ``Do \emph{all} the $p_i$ differ from zero at the same time?" Let us call the associated hypothesis $H_{1 2 \ldots N_T}$, which is to be compared with the GR hypothesis $\hyp_{\rm GR}$. 

A more interesting (because much more general) question would be: ``Do \emph{one or more} of the $p_i$ differ from zero", without specifying which. As we shall see, this question is more difficult to cast into the language of model selection. Below we will call the associated hypothesis $\hyp_{\rm modGR}$, to be compared with the GR hypothesis $\hyp_{\rm GR}$. What we will show is that, although there is no single waveform model that corresponds to $\hyp_{\rm modGR}$, testing the latter amounts to testing $2^{N_T} - 1$ hypotheses $H_{i_1 i_2 \ldots i_k}$ corresponding to all subsets $\{p_{i_1}, p_{i_2}, \ldots, p_{i_k}\}$ of the full set $\{p_1, p_2, \ldots, p_{N_T}\}$. Each of the hypotheses $H_{i_1 i_2 \ldots i_k}$ is tested by a waveform model in which $p_{i_1}, p_{i_2}, \ldots, p_{i_k}$ are free, but all the other $p_j$ are fixed to zero. The Bayes factors against GR for all of these tests can then be combined into a \emph{single} odds ratio which compares the full hypothesis $\hyp_{\rm modGR}$ with $\hyp_{\rm GR}$.

In a scenario with low signal-to-noise ratio (SNR), as will be the case with advanced ground-based detectors, parameter estimation will degrade significantly when trying to estimate too many parameters at once, and so will model selection if the alternative model to GR has too many additional degrees of freedom \cite{dpvv11}. This could be problematic if one only tests the `all-inclusive' hypothesis $H_{1 2 \ldots N_T}$ against GR, \emph{i.e.}, if the question one asks is ``Do all the $p_i$ differ from zero at the same time". The question we want to ask instead, namely ``Do one or more of the $p_i$ differ from zero", is not only more general; most of the sub-hypotheses $H_{i_1 i_2 \ldots i_k}$ involve a smaller number of free parameters, making the corresponding test more powerful in a low-SNR situation. And, as in most Bayesian frameworks, information from multiple sources can be combined in a straightforward way.


Our framework can be used with any family of waveforms with parameterized deformations, including the ppE family. In order to illustrate the method, following \cite{aiqs06a,aiqs06b,mais10} we make the simplest choice by adopting model waveforms in which the deviations away from GR take the form of shifts in a subset of the post-Newtonian inspiral phase coefficients. To establish the validity of the framework, we perform simulations with simple analytic frequency domain waveforms in stationary, Gaussian noise that follows the expected Advanced LIGO/Virgo noise curves. In the future, for actual tests of GR, one may want to use time domain waveforms and introduce deviations in \emph{e.g.} a Hamiltonian used to numerically evolve the motion of the binary.

As noted in \cite{yp09,dpvv11} and also illustrated here, if one is only interested in \emph{detection}, then it might suffice to only search with template waveforms predicted by GR, but parameter estimation can be badly off; this is what is called `fundamental bias'. Indeed, even if there is a deviation from GR in the signal, then a model waveform with completely different values for the masses and other parameters could still be a good fit to the data, with minimal loss of signal-to-noise ratio. We note that given model waveforms that feature a certain family of deformations away from GR, `fundamental bias' can still occur if the signal has a deviation that does not belong to the particular class of deformations allowed for by the models. However, in that case one expects the non-GR model waveforms to still be preferred over GR ones in the sense of \emph{model selection}, even if parameter estimation may be deceptive in interpreting the nature of the deviation. We will show an explicit example of this, and will argue that \emph{generic} deviations from GR can be picked up, of course subject to the limitations imposed by the detectors, as is the case with any kind of measurement. In particular, we expect that a GR violation will generally be visible, on condition that its effect on the phase at frequencies where the detector is the most sensitive is comparable to the effect of a few percent shift in one of the lower-order phase coefficients.

This paper is structured as follows. We first introduce our waveform model for compact binary inspiral, and discuss its sensitivity to changes in phase parameters (Section \ref{s:waveform}). In Section \ref{s:method}, we explain the basic method for single and multiple sources. In Section \ref{s:results}, we construct different simulated catalogs of sources and evaluate the level at which deviations from GR can be found. We end with a summary and a discussion of future steps to be taken.

Unless stated otherwise, we will take $G = c = 1$.

\section{Waveform model and its sensitivity to changes in phase coefficients}
\label{s:waveform}

We now introduce our waveform model. Since we are concerned with testing the strong-field dynamics of gravity, eventually all of the effects which we expect to see with compact binary coalescence should be represented in the waveform. This includes, but is not limited to, precession due to spin-orbit and spin-spin interactions \cite{bdiww95}, sub-dominant signal harmonics \cite{bfis08}, and merger/ringdown \cite{buonanno11}. In due time these will indeed need to be taken into account. However, in this paper we first and foremost wish to demonstrate the validity of a particular method, for which it will not be necessary to use very sophisticated waveforms. Also, as suggested by Fisher matrix calculations such as those of Mishra \textit{et al.}~\cite{mais10}, methods based on measuring phase coefficients will be the most accurate at low total mass. In this paper we limit ourselves to BNS sources, for which spin will be negligible, as well as sub-dominant signal harmonics \cite{vdbs06,vdbs07}. Since we will assume a network of Advanced LIGO and Virgo detectors, the merger and ringdown signals will also not have a large impact \cite{vz10}. Thus, for a first analysis we will focus on the inspiral part of the coalescence process, modeling the waveform in the frequency domain using the
stationary phase approximation (SPA) \cite{thorne87,sd91}. In particular, we use the so-called restricted TaylorF2 waveforms \cite{LAL,biops09} up to 2PN in phase.

Since the way we illustrate our method here is based on allowing for deviations in phase coefficients, we will need to know how sensitive our waveform model is to minor changes in the values of these coefficients. To get a sense of this, one could use the results of \cite{mais10} as a guide, but since these are based on the Fisher matrix they necessarily assume that signal and template are from the same waveform family. Before explaining our method for testing GR, we will first look at what happens both to detectability and parameter estimation when the signal contains a deviation from GR but is being searched for with a bank of GR templates.

\subsection{Model waveform(s) and detector configuration}

We start from the way TaylorF2 is implemented in the LIGO Algorithms Library \cite{LAL}:
\be
h(f) = \frac{1}{D} \frac{\mathcal{A}(\theta,\phi,\iota,\psi,\mathcal{M},\eta)}{\sqrt{\dot{F}(\mathcal{M},\eta;f)}} f^{2/3}\,e^{i\Psi(t_c, \phi_c, \mathcal{M},\eta;f)},
\label{TaylorF2}
\ee
where $D$ is the luminosity distance to the source, $(\theta,\phi)$ specify the sky position, $(\iota,\psi)$ give the orientation of the inspiral plane with respect to the line of sight, $\Mc$ is the chirp mass, and $\eta$ is the symmetric mass ratio. In terms of the component masses $(m_1,m_2)$, one has $\eta = m_1 m_2/(m_1 + m_2)^2$ and $\Mc = (m_1 + m_2)\,\eta^{3/5}$. $t_c$ and $\phi_c$ are the time and phase at coalescence, respectively. The `frequency sweep' $\dot{F}(\mathcal{M},\eta;f)$ is an expansion in powers of the frequency $f$ with mass-dependent coefficients, and
\ba
\label{psiidef}
&&\Psi(t_c, \phi_c, \mathcal{M},\eta;f) \nn\\
&&= 2\pi f t_c - \phi_c - \pi/4 \nn\\
&&\,\,\,\,\,\,\, + \sum_{i=0}^7 \left[\psi_i + \psi_{i}^{(l)} \ln f \right]\,f^{(i-5)/3}.
\label{phase}
\ea
In the case of GR, the functional dependence of the coefficients $\psi_i$ and $\psi_{i}^{(l)}$ on $(\mathcal{M},\eta)$ can be found in \cite{mais10}. However, here we will not assume that those relationships necessarily hold, except in the case of $\psi_0$, which has been tested using binary pulsars. With minor abuse of notation, let us re-label the remaining coefficients as $\psi_i$, $i = 1, \ldots, M$.

We note that $\dot{F}$ is related to the phase $\Psi$, and in principle we should also leave open the possibility that its expansion coefficients deviate from their GR values. However, with the Advanced LIGO and Virgo network and for stellar mass binaries we do not expect to be very sensitive to sub-dominant contributions to the amplitude \cite{vdbs06,vdbs07}.

Let us focus on the phase (\ref{phase}). One way of testing GR would be to use a model waveform in which
all the $\psi_i$ are considered free parameters, measure these together
with $\mathcal{M}$ and $\eta$, and check whether one obtains agreement with the functional relations
$\psi_i(\mathcal{M},\eta)$ predicted by GR. However, the events we expect in
Advanced LIGO/Virgo will probably not have sufficient SNR for this to be directly feasible \cite{mais10}.

Below, we instead suggest a scheme where a large number of tests are done, in each of which a specific, \emph{limited} subset
of the phase coefficients is left free while the others have the dependence on masses as in GR. The results
from all of these tests can then be combined into the odds ratio for a general deviation from GR versus GR being correct.

\begin{figure}[h]
\centering
\includegraphics[angle=0,width=\columnwidth]{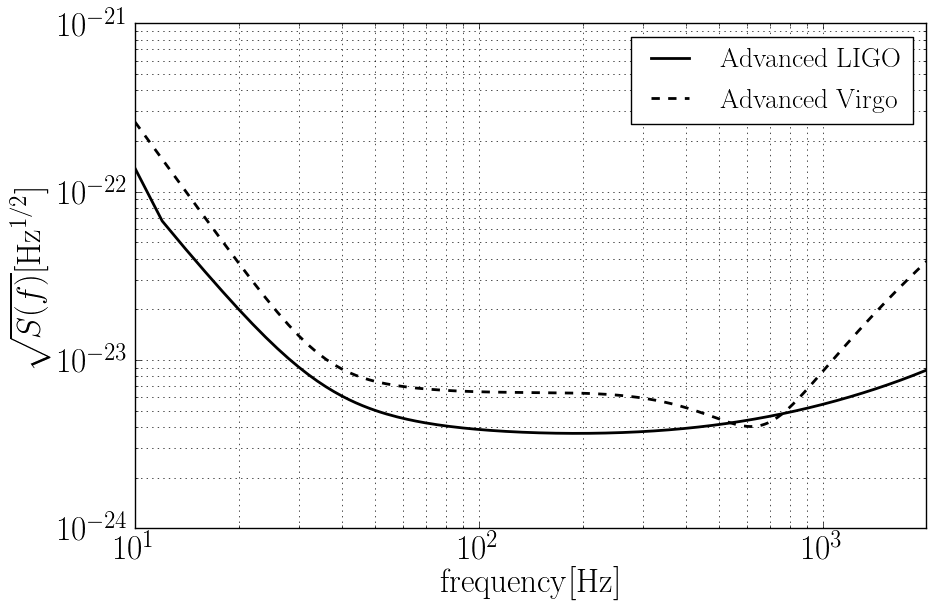}
\caption{The high-power, zero-detuning noise curve for Advanced LIGO, and the BNS-optimized Advanced Virgo noise curve.}
\label{fig:noisecurves}
\end{figure}

In this study we will assume a network of two Advanced LIGO detectors, one in Hanford, WA, and the other one in Livingston, LA, together with the Advanced Virgo detector in Cascina, Italy. We take the Advanced LIGO noise curve to be the one with zero-detuning of the signal recycling mirror and high laser power \cite{AdvLIGOnoise}. With these assumptions, the curves in Fig.~\ref{fig:noisecurves} represent the incoherent sums of the principal noise sources as they are currently understood; however, there may be unexpected, additional sources of noise. The high-power, zero-detuning option gives most of the desired sensitivity with the fewest technical difficulties. Advanced Virgo can also be optimized for BNS sources by an appropriate choice of the signal recycling detuning and the signal recycling mirror transmittance \cite{aVirgo}, and this is what we assume here.

\subsection{Changes in phase coefficients and detectability}

It is  important to investigate to what extent signals that may deviate from General Relativity could be detected in the first place. The fitting factor ($FF$) is a measure of the adequateness of a template family to fit the signal; $1-FF$ is the reduction in signal-to-noise ratio that occurs from using a model waveform which differs from the exact signal waveform when searching the data. Let $h_e(\vec{\lambda})$ be the `exact' waveform of the signal and $h_m(\vec{\theta})$
the model used for detection; the exact and model waveforms are dependent on sets of parameters $\vec{\lambda}$ and $\vec{\theta}$, respectively.
The fitting factor is then defined as \cite{apostolatos95}
\begin{equation}
FF \equiv \max_{\vec{\theta}} \left(\frac{\langle h_e(\vec{\lambda}) \mid h_m(\vec{\theta})\rangle}{\sqrt{\langle h_m(\vec{\theta})\mid h_m (\vec{\theta} ) \rangle \langle h_e(\vec{\lambda}) \mid h_e(\vec{\lambda}) \rangle }}\right)\,,
\end{equation}
where $\langle a \mid b \rangle$ denotes the usual noise weighted inner product,
\begin{equation}
 \langle a \mid b \rangle = 2\int^{f_\mathrm{max}}_{f_\mathrm{min}} \frac{a^*(f)b(f)+a(f)b^*(f)}{S_{n}(f)} df\,,
\end{equation}
and $S_n(f)$ is the one-sided noise spectral density \cite{Maggiore}. $f_{\rm min}$ is the detectors' lower cut-off frequency, and in our case, $f_{\rm max}$ is the frequency at last stable orbit, $f_{\rm max} = (6^{3/2} \pi \mathcal{M} \eta^{-3/5})^{-1}$.
We note that the detection rate scales like the cube of the signal-to-noise ratio, so that the fractional reduction in event rate is $1-FF^3$ \cite{apostolatos95}. In this case the waveform is not `exact' in the sense of numerical relativity; instead we use the fitting factor as a measure of how similar a modified TaylorF2 waveform and a GR version are.

\begin{figure}[h]
\centering
\includegraphics[angle=0,width=\columnwidth]{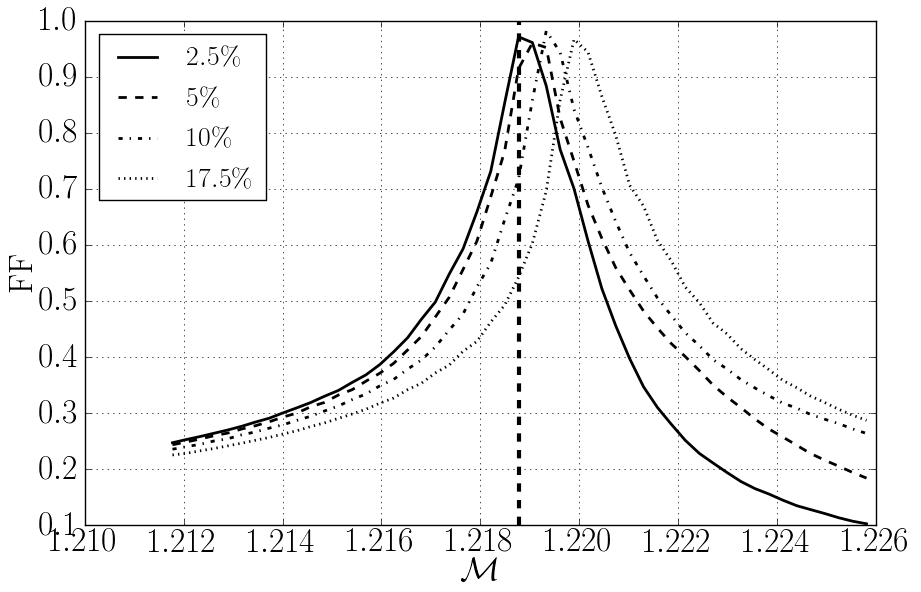}
\caption{The fitting factors for a range of $\Mc$, once the other parameters are maximized over. Here a deviation of $\psi_3$ from $2.5\%$ to $17.5\%$ is used for a  $(1.4,1.4)\,M_\odot$ system. The vertical dashed line represents the true value of $\Mc$, while the maximum is offset to compensate for the modification in the phase.}
\label{f:mchirp delta psi3 =17.5}
\end{figure}

\begin{figure}[h]
\centering
\includegraphics[angle=0,width=\columnwidth]{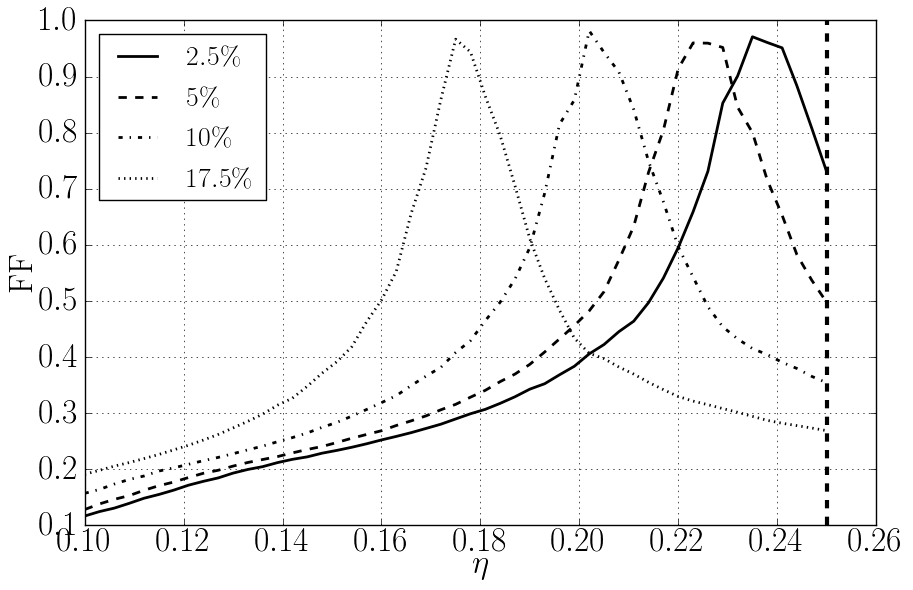}
\caption{The fitting factors for a range of $\eta$, once the other parameters are maximized over. Here a deviation of $\psi_3$ from $2.5\%$ to $17.5\%$ is used for a  $(1.4,1.4)\,M_\odot$ system. The vertical dashed line represents the true value of $\eta$, while the maximum is offset to compensate for the modification in the phase.}
\label{f:eta delta psi3 =17.5}
\end{figure}

A sample deviation is tested using a modified waveform different only in the 1.5PN order phase coefficient: $\psi_3^{\rm GR}(\Mc,\eta) \rightarrow  \psi_3^{\rm GR}(\Mc,\eta)\,\left[1 + \delta\chi_3 \right]$. The Advanced LIGO noise curve is used. The signal is `detected' with a template bank
of standard GR waveforms that is regularly spaced in the parameters present in the phase, $\phi_c$, $t_c$, $\mathcal{M}$, and $\eta$.
We study a $(1.4, 1.4)\,M_\odot$ binary with $\delta\chi_3$ ranging from 0.025 to 0.175. As seen in figures \ref{f:mchirp delta psi3 =17.5} and \ref{f:eta delta psi3 =17.5}, the mass parameters can absorb the change in the phase due to the modified phase coefficient while providing a fitting factor of over 95\%.

Thus, with a template bank of GR waveforms it is possible to detect a signal containing a large deviation from GR without significant loss in signal-to-noise ratio, but recovered with intrinsic parameters that deviate significantly from the true values.
We now describe a method which will be able to nevertheless recognize a deviation from GR when one is present.

\section{Method}
\label{s:method}

We will first recall some basic facts about Bayesian inference. Next we outline our method for finding a possible violation of GR using inspiral signals by allowing for deviations in phase coefficients. For simplicity, we start with an example where only two phase coefficients are taken into account; then we go on to the general case. Finally we explain how to combine information from a catalog of sources.

\subsection{Bayesian inference and nested sampling}

We now give a brief overview of Bayesian inference, as well as the method of nested sampling due to Skilling \cite{Skilling:AIP}, which was first introduced into ground-based GW data analysis by Veitch and Vecchio \cite{VeitchVecchio:2008a,VeitchVecchio:2008b,vv09} and which we will adopt here as well.

Let us consider hypotheses $\hyp_i$, $\hyp_j$. Here $\hyp_i$ could be the hypothesis that there is a deviation from GR while $\hyp_j$ is the hypothesis that GR is correct; or $\hyp_i$ could simply be the hypothesis that a signal of a particular form is present in the data while $\hyp_j$ is the hypothesis that there is only noise. The statements that we can make about any hypothesis are based on a data set $d$ (observations) and all the relevant prior information $\info$ that we hold.

Within the framework of Bayesian inference, the key quantity that one needs to compute is the posterior probability of a hypothesis $\hyp_i$. Applying Bayes' theorem we obtain
\be
P(\hyp_i | d, \info) = \frac{P(\hyp_i | \info)\,P(d |\hyp_i, \info)}{P(d | \info)}\,,
\label{posterior}
\ee
where $P(\hyp_i | d, \info)$ is the \emph{posterior probability} of the hypothesis $\hyp_i$ given the data, $P(\hyp_i | \info)$ is the \emph{prior probability} of the hypothesis, and $P(d |\hyp_i, \info)$ is the \emph{marginal likelihood} or \emph{evidence} for $\hyp_i$, which can be written as:
\ba
P(d |\hyp_i, \info) & = & {\cal L}({\cal H}_i)
\nonumber\\
& = & \int d\vec{\theta}\, p(\vec{\theta} | {\cal H}_i, \info)\, p(d | \vec{\theta}, {\cal H}_i, \info)\,.
\label{e:marglikelihood}
\ea
In the above expression, $p(\vec{\theta} | {\cal H}_i, \info)$ is the prior probability density of the unknown parameter vector $\vec{\theta}$ within the model corresponding to ${\cal H}_i$, and $p(d | \vec{\theta}, {\cal H}_i, \info)$ is the likelihood function of the observation $d$, assuming a given value of the parameters $\vec{\theta}$ and the model ${\cal H}_i$.

If we want to compare different hypotheses, $\hyp_i$ and $\hyp_j$, in light of the observations made, we can compute the ratio of posterior probabilities, which is known as the \emph{odds ratio}:
\ba\label{e:bayes}
O^i_j & = & \frac{P({\cal H}_i|d, \info)}{P({\cal H}_j|d, \info)}
\nonumber\\
& = & \frac{P({\cal H}_i|\info)}{P({\cal H}_j|\info)}\frac{P(d | {\cal H}_i, \info)}{P(d | {\cal H}_j, \info)}
\nonumber\\
& = &  \frac{P({\cal H}_i|\info)}{P({\cal H}_j|\info)}\,B^i_j\,,
\ea
where $P({\cal H}_j|\info)/P({\cal H}_i|\info)$ is the \emph{prior odds} of the two hypotheses, the relative confidence we assign to the models before any observation, and $B^i_j$ is the \emph{Bayes factor}.

In addition to computing the relative probabilities of different models, one usually wants to make inference on the unknown parameters, and therefore one needs to compute the joint posterior probability density function (PDF)
\be
p(\vec{\theta} | d, {\cal H}_i, \info) = \frac{p(\vec{\theta} | {\cal H}_i, \info) p(d | \vec{\theta}, {\cal H}_i, \info)}{p(d|{\cal H}_i, \info)}\,.
\label{e:pdf0}
\ee
From the previous expression it is simple to compute the marginalized PDF on any given parameter, say $\theta_1$ within a given model ${\cal H}_i$:
\be
p(\theta_1 | d, {\cal H}_i, \info) = \int d\theta_2 \dots \int d\theta_N\, p(\vec{\theta} | d, {\cal H}_i, \info)\,.
\label{e:pdf}
\ee

The key quantities for Bayesian inference in Eq.~(\ref{e:bayes}),~(\ref{e:pdf0}) and~(\ref{e:pdf}) can be efficiently computed using \emph{e.g.} a nested sampling algorithm~\cite{Skilling:AIP}. The basic idea of nested sampling is to use a collection of $n$ objects, called \emph{live points}, randomly sampled from the prior distributions, but subject to a constraint over the value of their likelihood. A live point $\vec{\xi}$ is a point in the multidimensional parameter space. At each iteration, the live point $\vec{\xi}^*$ having the lowest likelihood $\cal{L}(\vec{\xi}^*)$ is replaced with a new point $\vec{\xi}$ sampled from the prior distribution. To be accepted, the new point must obey to the condition
\begin{equation}
\cal{L}(\vec{\xi})>\cal{L}(\vec{\xi}^*)\,.
\end{equation}
The above condition ensures that regions of progressively increasing likelihood are explored, and the evidence integral, Eq.~(\ref{e:marglikelihood}), is calculated using those points as the computation progresses.

In this paper we use a specific implementation of this technique that was developed for ground-based observations of coalescing binaries by Veitch and Vecchio; we point the interested reader to \cite{VeitchVecchio:2008a,VeitchVecchio:2008b,vv09,AylottEtAl:2009,AVV:2009} for technical details. To select a new live point, a Metropolis-Hastings Markov chain Monte Carlo (MCMC) is used with $p$ steps. The uncertainty in the evidence computation for a given number of live points $n$ and MCMC steps $p$ was quantified in \cite{vv09}. For the calculations in this paper we took $n = 1000$ and $p = 100$, in which case the standard deviation on log Bayes factors is $\mathcal{O}(1)$.

\subsection{Basic method for a single source}

Given that we have no knowledge of which coefficient(s) might deviate from the GR values, we want to test the hypothesis that at least
one of the known coefficients $\{\psi_0, \psi_1, \psi_2, \ldots, \psi_M\}$ is different. Computational limitations and possible lack of sensitivity to changes in the higher-order coefficients will induce us to only look for deviations in a set of \emph{testing coefficients} $\{\psi_1, \ldots, \psi_{N_T}\} \subset \{\psi_0, \psi_1, \ldots, \psi_M\}$, with $N_T < M$. For the examples of Sec.~\ref{s:results}, we will choose $N_T = 3$ due to computational constraints, but a larger number could be used.

Let us introduce some notation. We define hypotheses $H_{i_1 i_2 \ldots i_k}$ as follows:
\begin{quote}
$H_{i_1 i_2 \ldots i_k}$ is the hypothesis that the phasing coefficients $\psi_{i_1}, \ldots, \psi_{i_k}$ do not have the functional dependence on $(\Mc,\eta)$ as predicted by General Relativity, but all other coefficients $\psi_j$, $j \notin \{i_1, i_2, \ldots, i_k\}$ \emph{do} have the dependence as in GR.
\end{quote}
Thus, for example, $H_{12}$ is the hypothesis that $\psi_1$ and $\psi_2$ deviate from their GR values, with all other coefficients being as in GR. With each of
the hypotheses above, we can associate a waveform model that can be used to test it. Let $\vec{\theta} = \{\Mc, \eta, \ldots\}$ be the parameters occurring in the GR waveform. Then $H_{i_1 i_2 \ldots i_k}$ is tested by a waveform in which the independent parameters are
\be
\{\vec{\theta}, \psi_{i_1}, \psi_{i_2}, \ldots, \psi_{i_k} \},
\ee
\emph{i.e.} the coefficients $\{\psi_{i_1}, \psi_{i_2} \ldots, \psi_{i_k}\}$ are allowed to vary freely. In practice, these will need to be subsets of a limited set of coefficients; in Section \ref{s:results}, where we present results, we will consider all subsets of the set $\{\psi_1, \psi_2, \psi_3\}$. This choice will already suffice to illustrate the method without being overly computationally costly, but in the future one may want to use a larger set.

Now, the hypothesis we would really like to test is that one or more of the $\psi_i$ differ from GR, without specifying which. This corresponds to a logical `or' of the above hypotheses:
\be
\hyp_{\rm modGR} = \bigvee_{i_1 < i_2 < \ldots < i_k} H_{i_1 i_2 \ldots i_k}.
\label{logicalOR}
\ee
Our aim is to compute the following \emph{odds ratio}:
\be
O^{\rm modGR}_{\rm GR} \equiv \frac{P(\hyp_{\rm modGR}|d,\info)}{P(\hyp_{\rm GR}|d,\info)}.
\ee
However, the hypothesis (\ref{logicalOR}) is not what model waveforms with one or more free coefficients $\psi_i$ will test; rather, such waveforms test the hypotheses $H_{i_1 i_2 \ldots i_k}$ themselves. What we will need to do is to break up the logical `or' in $\hyp_{\rm modGR}$ into the component hypotheses $H_{i_1 i_2 \ldots i_k}$. Fortunately, this is trivial.

Before treating the problem more generally, let us consider a simple example.
Imagine that only two coefficients $\psi_1$ and $\psi_2$ are being used for the testing of GR. Then
\be
\hyp_{\rm modGR} = H_1 \vee H_2 \vee H_{12}.
\label{h1orh2}
\ee
In this example, the odds ratio of interest can then be written as
\be
{}^{(2)}O^{\rm modGR}_{\rm GR} \equiv \frac{P(H_1 \vee H_2 \vee H_{12}|d,\info)}{P(\hyp_{\rm GR}|d,\info)},
\ee
where the superscript $(2)$ reminds us that only two of the parameters are being used for testing.

An important observation is that the hypotheses $H_1$, $H_2$, $H_{12}$ are \emph{logically disjoint}: the `and' of any two of them is always false. Indeed,
in $H_1$, $\psi_2$ takes the value predicted by GR, but in $H_2$ it differs from the GR value, as it does in $H_{12}$.
Similarly, in $H_2$, $\psi_1$ takes the GR value, but in $H_1$ it differs from the GR value, and the same in $H_{12}$. More generally, any two hypotheses
$H_{i_1 i_2 \ldots i_k}$ and $H_{j_1 j_2 \ldots j_l}$ with $\{i_1, i_2, \ldots, i_k\} \neq \{j_1, j_2, \ldots, j_l\}$ are logically disjoint. This means that the odds
ratio is simply
\begin{widetext}
\be
{}^{(2)}O^{\rm modGR}_{\rm GR} = \frac{P(H_1|d,\info)}{P(\hyp_{\rm GR}|d,\info)} + \frac{P(H_2|d,\info)}{P(\hyp_{\rm GR}|d,\info)} + \frac{P(H_{12}|d,\info)}{P(\hyp_{\rm GR}|d,\info)}.
\ee
\end{widetext}
Using Bayes' theorem, this can be written as
\begin{widetext}
\be
{}^{(2)}O^{\rm modGR}_{\rm GR} = \frac{P(H_1|\info)}{P(\hyp_{\rm GR}|\info)}\,B^1_{\rm GR} + \frac{P(H_2|\info)}{P(\hyp_{\rm GR}|\info)}\,B^2_{\rm GR}
+ \frac{P(H_{12}|\info)}{P(\hyp_{\rm GR}|\info)}\,B^{12}_{\rm GR}.
\label{Oddsfinal}
\ee
\end{widetext}
Here $B^1_{\rm GR}$, $B^2_{\rm GR}$, $B^{12}_{\rm GR}$ are the Bayes factors
\ba
B^1_{\rm GR} &=& \frac{P(d|H_1, \info)}{P(d| \hyp_{\rm GR}, \info)}, \nn\\
B^2_{\rm GR} &=& \frac{P(d|H_2, \info)}{P(d| \hyp_{\rm GR}, \info)}, \nn\\
B^{12}_{\rm GR} &=& \frac{P(d|H_{12}, \info)}{P(d| \hyp_{\rm GR}, \info)},
\label{Bs}
\ea
and $P(H_1|\info)/P(\hyp_{\rm GR}|\info)$, $P(H_2|\info)/P(\hyp_{\rm GR}|\info)$, $P(H_{12}|\info)/P(\hyp_{\rm GR}|\info)$
are ratios of prior odds.

In practice, we will write the testing coefficients as
\be
\psi_i = \psi_i^{\rm GR}(\mathcal{M},\eta)\,\left[1 + \delta\chi_i\right],
\label{deltapsi}
\ee
with $\psi_i^{\rm GR}(\Mc,\eta)$ the functional form of the dependence of $\psi_i$ on $(\Mc,\eta)$ according to GR, and the dimensionless $\delta\chi_i$ is a fractional shift in $\psi_i$. Note that in GR, the 0.5PN contribution is identically zero; it will be treated separately, as explained in section \ref{s:results}. In the example of this subsection, one can assume that $\psi_1$, $\psi_2$ are any of the PN coefficients \emph{other} than the 0.5PN one.
With the above notation, the Bayes factors (\ref{Bs}) are
\begin{widetext}
\ba
B^1_{\rm GR}
    &=& \frac{\int d\vec{\theta}\, d\delta\chi_1 \,\,\,{}^{\{1\}}\pi(\delta\chi_1)\, \pi(\vec{\theta})\, p(d|\vec{\theta}, \delta\chi_1, H_1, \info) }{\int d\vec{\theta} \,\pi(\vec{\theta}) \,p(d|\vec{\theta},\hyp_{\rm GR},\info)} , \label{newBayesfactor1}\\
B^2_{\rm GR}
    &=& \frac{\int d\vec{\theta}\, d\delta\chi_2 \,\,\,{}^{\{2\}}\pi(\delta\chi_2)\, \pi(\vec{\theta})\, p(d|\vec{\theta}, \delta\chi_2, H_2, \info) }{\int d\vec{\theta}\, \pi(\vec{\theta})\, p(d|\vec{\theta},\hyp_{\rm GR},\info)} , \label{newBayesfactor2}\\
B^{12}_{\rm GR}
       &=& \frac{\int d\vec{\theta} \,d\delta\chi_1\, d\delta\chi_2 \,\,\,{}^{\{12\}}\pi(\delta\chi_1, \delta\chi_2)\, \pi(\vec{\theta})\, p(d|\vec{\theta}, \delta\chi_1, \delta\chi_2, H_{12}, \info) }{\int d\vec{\theta}\, \pi(\vec{\theta})\, p(d|\vec{\theta},\hyp_{\rm GR},\info)} .
\label{newBayesfactor3}
\ea
\end{widetext}
Here ${}^{\{1\}}\pi(\delta\chi_1)$, ${}^{\{2\}}\pi(\delta\chi_2)$, ${}^{\{12\}}\pi(\delta\chi_1, \delta\chi_2)$ are priors for, respectively, $\delta\chi_1$, $\delta\chi_2$, and the pair $(\delta\chi_1, \delta\chi_2)$. We choose these to be constant functions in the relevant parameter or pair of parameters, having support within a large interval or square centered on the origin, and normalized to one. For the other parameters, $\vec{\theta}$, we use the same functional form and limits as \cite{vv09}, with the exception of the distance being allowed to vary between $1$ and $1000$\,Mpc.  Specifically, for the sky location and the orientation of the orbital plane we choose uniform priors on the corresponding unit spheres. For the phase at coalescence $\phi_c$ we choose a flat prior with $\phi_c \in [0, 2\pi]$, and the time of coalescence $t_c$ is in a time interval of 100 ms. The prior on $\eta$ is flat on the interval $[0, 0.25]$. For chirp mass we use an approximation to the Jeffreys prior which gives $p(\Mc|\info) \propto \Mc^{-11/6}$; see \cite{vv09} for motivation. In addition, component masses are restricted to the interval $m_1, m_2 \in [1, 34]\,M_\odot$.


At this point it is worth commenting on the mutual relationships of the hypotheses $H_{i_1 i_2 \ldots i_k}$ amongst each other and with $\hyp_{\rm GR}$, and on the waveform models used to test them. As an example, let us discuss the case of $H_1$ and $\hyp_{\rm GR}$. Consider the numerator of the Bayes factor $B^1_{\rm GR}$ in Eq.~(\ref{newBayesfactor1}):
\be
\int d\vec{\theta}\, d\delta\chi_1 \,\,\,{}^{\{1\}}\pi(\delta\chi_1)\, \pi(\vec{\theta})\, p(d|\vec{\theta}, \delta\chi_1, H_1, \info).
\ee
The parameter space of the GR waveforms, $\{\vec{\theta}\}$, has a natural embedding into the parameter space $\{\vec{\theta}, \delta\chi_1\}$ of the waveforms used to test $H_1$: it can be identified with the hypersurface $\delta\chi_1 = 0$. We could have explicitly excluded this hypersurface from $\{\vec{\theta},\delta\chi_1\}$ by setting a prior on $\delta\chi_1$ of the form ${}^{\{1\}}\pi_0(\delta\chi_1) = 0$ if $\delta\chi_1 = 0$ and ${}^{\{1\}}\pi_0(\delta\chi_1) = \mbox{const}$ otherwise. However, this would not have made a difference in the integral above; indeed, with respect to the integration measure induced by the prior probability density on $\{\vec{\theta}, \delta\chi_1\}$, the surface $\delta\chi_1 = 0$ constitutes a set of measure zero anyway. Now look at the denominator in the expression for $B^1_{\rm GR}$, which is the evidence for the GR hypothesis:
\be
\int d\vec{\theta}\, \pi(\vec{\theta})\, p(d|\vec{\theta},\hyp_{\rm GR},\info).
\ee
Despite the fact that the GR waveforms form a set of measure zero within the set of waveforms used for testing $H_1$, the above integral is clearly not zero. It is the evidence for a qualitatively different hypothesis, whose parameter space $\{\vec{\theta}\}$ carries a different integration measure with respect to which the marginalization of the likelihood is carried out. In particular, $\hyp_{\rm GR}$ is not `included' in the $H_1$ model in any sense that is meaningful for model selection.

Next let us consider $H_{12}$. The numerator of the Bayes factor $B^{12}_{\rm GR}$ in Eq.~(\ref{newBayesfactor3}) is
\be
\int d\vec{\theta} \,d\delta\chi_1\, d\delta\chi_2 \,\,\,{}^{\{12\}}\pi(\delta\chi_1, \delta\chi_2)\, \pi(\vec{\theta})\, p(d|\vec{\theta}, \delta\chi_1, \delta\chi_2, H_{12}, \info).
\ee
The parameter space of the GR waveforms has a natural embedding into the parameter space $\{\vec{\theta}, \delta\chi_1, \delta\chi_2\}$ of the waveforms used to test $H_{12}$, by identification with the hypersurface $\delta\chi_1 = \delta\chi_2 = 0$. Similarly, the parameter spaces of the waveforms we use to test $H_1$ and $H_2$ can be identified with the hypersurface $\delta\chi_2 = 0$ and the hypersurface $\delta\chi_1 = 0$, respectively. With respect to the integration measure induced by the prior on $\{\vec{\theta}, \delta\chi_1, \delta\chi_2\}$, these hypersurfaces have measure zero. Despite this, the numerators of $B^1_{\rm GR}$ and $B^2_{\rm GR}$ in Eqns.~(\ref{newBayesfactor1}) and (\ref{newBayesfactor2}) are not zero, because of the different integration measures. With our choices of prior probability densities on the different parameter spaces, and the associated waveform models, there is no meaningful sense in which $H_1$ or $H_2$ are `included' in $H_{12}$. Thus, we are indeed testing the disjoint hypotheses $H_1$, $H_2$, and $H_{12}$. The latter is the hypothesis that \emph{both} $\delta\chi_1$ \emph{and} $\delta\chi_2$ differ from zero: the prior density ${}^{\{12\}}\pi(\delta\chi_1, \delta\chi_2)$ gives the surfaces $\delta\chi_1 = 0$ and $\delta\chi_2 = 0$ zero prior mass.

The waveform model that leaves $\{\vec{\theta}, \delta\chi_1, \delta\chi_2\}$ free corresponds to the hypothesis $H_{12}$. When computing the Bayes factor $B^{12}_{\rm GR}$, the question one addresses is ``Do $\psi_1$ \emph{and} $\psi_2$ both differ from what GR predicts?" This is analogous to what has been done in recent Bayesian work on testing GR \cite{dpvv11,csyp11,gvs11}. As we are in the process of showing, it is possible to address a more general question, namely ``Do $\psi_1$, \emph{or} $\psi_2$, \emph{or} both at the same time, differ from their GR values?"

For completeness, we note that what nested sampling will give us directly are not the Bayes factors of Eqns.~(\ref{newBayesfactor1})--(\ref{newBayesfactor3}), but rather the Bayes factors for the various hypotheses against the noise-only hypothesis $\hyp_{\rm noise}$:
\ba
B^1_{\rm noise} &=& \frac{P(d|H_1, \info)}{P(d| \hyp_{\rm noise},\info)}, \nn\\
B^2_{\rm noise} &=& \frac{P(d|H_2, \info)}{P(d| \hyp_{\rm noise},\info)}, \nn\\
B^{12}_{\rm noise} &=& \frac{P(d|H_{12}, \info)}{P(d| \hyp_{\rm noise},\info)}, \nn\\
B^{\rm GR}_{\rm noise} &=& \frac{P(d|\hyp_{\rm GR}, \info)}{P(d| \hyp_{\rm noise},\info)}.
\ea
These can trivially be combined to obtain the Bayes factors of Eqns.~(\ref{newBayesfactor1})--(\ref{newBayesfactor3}):
\be
B^1_{\rm GR} = \frac{B^1_{\rm noise}}{B^{\rm GR}_{\rm noise}},\,\,\,\,\,B^2_{\rm GR} = \frac{B^2_{\rm noise}}{B^{\rm GR}_{\rm noise}},\,\,\,\,\,B^{12}_{\rm GR} = \frac{B^{12}_{\rm noise}}{B^{\rm GR}_{\rm noise}}.
\ee
Now, upon calculating the Bayes factors for each model, we would like to combine these measurements into an overall odds ratio between the GR model and \emph{any} of the competing hypotheses (Eq.~(\ref{Oddsfinal})). In order to do this, we must specify the prior odds for each model against GR, $P(H_1|\info)/P(\hyp_{\rm GR}|\info)$, $P(H_2|\info)/P(\hyp_{\rm GR}|\info)$, \emph{etc.} Here one might want to let oneself be guided by, \emph{e.g.}, the expectation that a violation of GR will likely occur at higher post-Newtonian order, and give more weight to $H_2$ and $H_{12}$. Or, if one expects a deviation to happen only in a particular phase coefficient (such as $\psi_2$ in the case of `massive gravity'), one may want to downweight the most inclusive hypothesis, in this example $H_{12}$. In reality, we will not know beforehand what form a violation will take; in particular, it could affect all of the PN coefficients. For the purposes of this analysis, we invoke the principle of indifference among the alternative hypotheses, taking no one to be preferable to any other. This imposes the condition that the prior odds of each against GR are equal. We explicitly note that this is a choice of the authors for computing final results. This choice results in our effectively taking the average of the Bayes factors for the alternative hypotheses when we compute the odds ratio versus GR.

When combining the Bayes factors into the odds ratio, we therefore assume

\begin{equation}
\frac{P(H_1|\info)}{P(\hyp_{\rm GR}|\info)} = \frac{P(H_2|\info)}{P(\hyp_{\rm GR}|\info)} =
\frac{P(H_{12}|\info)}{P(\hyp_{\rm GR}|\info)}.\label{restrictedprioroddsratios}
\end{equation}

Furthermore, we let
\be
\frac{P(\hyp_{\rm modGR}|\info)}{P(\hyp_{\rm GR}|\info)} =
\frac{P(H_1 \vee H_2 \vee H_{12}|\info)}{P(\hyp_{\rm GR}|\info)} = \alpha,
\ee
where we do not specify $\alpha$; it will end up being an overall scaling of the odds ratio. This, together with (\ref{restrictedprioroddsratios}) and the logical disjointness of the hypotheses $H_1$, $H_2$, $H_{12}$ implies
\be
\frac{P(H_1|\info)}{P(\hyp_{\rm GR}|\info)} = \frac{P(H_2|\info)}{P(\hyp_{\rm GR}|\info)} = \frac{P(H_{12}|\info)}{P(\hyp_{\rm GR}|\info)} = \frac{\alpha}{3}.
\ee

The final expression for the odds ratio for a modification of GR versus GR, in the case where up to two coefficients are used for testing, is then
\be
{}^{(2)}O^{\rm modGR}_{\rm GR} = \frac{\alpha}{3} \left[B^1_{\rm GR} + B^2_{\rm GR} + B^{12}_{\rm GR}\right].
\label{logOdds2params}
\ee

Before continuing to the case of more than two testing parameters, let us compare and contrast what is proposed here with what was done in previous Bayesian work, \emph{e.g.} \cite{dpvv11,csyp11,gvs11}. There, one introduced free parameters $p_i$ in the waveform (not necessarily corresponding to shifts in phase coefficients; they could, \emph{e.g.}, be shifts in ringdown frequencies and damping times), which are zero in GR. Next, one constructed a Bayes factor comparing a model waveform in which all of the $p_i$ are allowed to vary freely with the GR model in which all of the $p_i$ are fixed to zero. In our language and in the case of two testing parameters, this corresponds to only comparing the hypothesis $H_{12}$ with the GR hypothesis $\hyp_{\rm GR}$. Hence, what was effectively done in previous work was to address the question: ``Do \emph{all} of the additional free parameters differ from zero at the same time?" A more interesting question to ask is ``Do \emph{one or more} of the extra parameters differ from zero?" This corresponds to testing our hypothesis $\hyp_{\rm modGR}$. There is no waveform model that can be used to test the latter hypothesis directly, but as we were able to show, $\hyp_{\rm modGR}$ can be broken up into sub-hypotheses, $H_1$, $H_2$, and $H_{12}$ in the case of two testing parameters. With each of these, a waveform model \emph{can} be associated, hence they can be tested against the GR hypothesis, $\hyp_{\rm GR}$. The resulting Bayes factors can be combined into an odds ratio as in Eq.~(\ref{logOdds2params}), which \emph{does} compare the more general hypothesis $\hyp_{\rm modGR}$ with $\hyp_{\rm GR}$.

\subsection{The general case}
\label{subsec:general}

So far we have assumed just two testing coefficients, but we may want to use more. In practice it makes sense to pick $\{\psi_1 ,\ldots, \psi_{N_T}\}$, $N_T \leq M$. The number of coefficients used, $N_T$, will be dictated mostly by computational cost; in Sec.~\ref{s:results} we will pick $N_T = 3$ but a larger number could be chosen. We then define
\be
\hyp_{\rm modGR} = \bigvee_{i_1 < i_2 < \ldots < i_k; k \leq N_T} H_{i_1 i_2 \ldots i_k}.
\ee
When using this set of testing coefficients, the odds ratio for `modification to GR' versus GR
becomes:
\ba
&&{}^{(N_T)}O^{\rm modGR}_{\rm GR} \nn\\
&&= \frac{P(\hyp_{\rm modGR}|d,\info)}{P(\hyp_{\rm GR}|d,\info)} \nn\\
&&= \frac{P(\bigvee_{i_1 < i_2 < \ldots < i_k; k \leq N_T} H_{i_1 i_2 \ldots i_k}|d,\info)}{P(\hyp_{\rm GR}|d,\info)},\nn\\
\ea
where as before, $H_{i_1 i_2 \ldots i_k}$ is the hypothesis that $\{\psi_{i_1}, \psi_{i_2}, \ldots, \psi_{i_k}\}$ do not have the functional dependence on $(\mathcal{M}, \eta)$ as predicted by GR, but all of the remaining coefficients do.
Thus, we are considering the odds ratio for one or more of the phase coefficients $\psi_1, \ldots, \psi_{N_T}$ deviating from GR, versus all of them
having the functional dependence on masses as in GR.

Using the logical disjointness of the $H_{i_1 i_2 \ldots i_k}$ for different subsets $\{i_1, i_2, \ldots, i_k\}$ as well as Bayes' theorem, one can write
\be
{}^{(N_T)}O^{\rm modGR}_{\rm GR} = \sum_{k=1}^{N_T} \sum_{i_1 < i_2 < \ldots < i_k} \frac{P(H_{i_1 i_2 \ldots i_k}|\info)}{P(\hyp_{\rm GR}|\info)} B^{i_1 i_2 \ldots i_k}_{\rm GR},
\ee
where
\be
B^{i_1 i_2 \ldots i_k}_{\rm GR} = \frac{P(d|H_{i_1 i_2 \ldots i_k},\info)}{P(d|\hyp_{\rm GR},\info)}.
\ee

Again, one computes the $2^{N_T}-1$ individual Bayes factors $B^{i_1 i_2 \ldots i_k}_{\rm GR}$ of each of the alternative hypotheses versus GR. The evaluation of the odds ratio requires that we use specific values for the prior odds ratios. We will set them equal to each other, as we did in Eq.~(\ref{restrictedprioroddsratios}):
\be
\frac{P(H_{i_1 i_2 \ldots i_k}|\info)}{P(\hyp_{\rm GR}|\info)} = \frac{P(H_{j_1 j_2 \ldots j_l}|\info)}{P(\hyp_{\rm GR}|\info)}\,\,\,\,\,\,\,\mbox{for any $k, l \leq N_T$},
\label{generalprioroddsratios}
\ee
in which case the odds ratio ${}^{(N_T)}O^{\rm modGR}_{\rm GR}$ will be proportional to a straightforward average of the Bayes factors. We note once again that other choices could in principle be made. If one expects a violation to be mainly visible in a \emph{particular} phase coefficient (see, \emph{e.g.}, Table I in \cite{csyp11}), then one might want to downweight the more inclusive hypotheses. Or, one may argue that a deviation will most likely affect \emph{all} of the coefficients starting from some PN order, but then it would not necessarily be sensible to give more inclusive hypotheses a lower weight. Hence we assign equal prior odds to all of the sub-hypotheses $H_{i_1 i_2 \ldots i_k}$.

Also as before, we let
\be
\frac{P(\hyp_{\rm modGR}|\info)}{P(\hyp_{\rm GR}|\info)} = \alpha,
\label{totalignorance}
\ee
where we do not specify $\alpha$; it will end up being an overall prefactor in the odds ratio.
The equality (\ref{totalignorance}), together with (\ref{generalprioroddsratios}) and the logical disjointness of the
$2^{N_T} - 1$ hypotheses $H_{i_1 \ldots i_k}$ implies
\be
\frac{P(H_{i_1 i_2 \ldots i_k}|\info)}{P(\hyp_{\rm GR}|\info)} = \frac{\alpha}{2^{N_T} - 1}.
\label{eq:component-odds}
\ee
In terms of the $H_{i_1 i_2 \ldots i_k}$, the odds ratio can then be written as
\be
{}^{(N_T)}O^{\rm modGR}_{\rm GR} = \frac{\alpha}{2^{N_T}-1}\,\sum_{k=1}^{N_T} \sum_{i_1 < i_2 < \ldots < i_k} B^{i_1 i_2 \ldots i_k}_{\rm GR}.
\label{oddsindividual}
\ee
Analogously to (\ref{newBayesfactor1})--(\ref{newBayesfactor3}), we will use priors on the parameters $\delta\chi_i$
\be
{}^{\{i_1 i_2 \ldots i_k\}}\pi(\delta\chi_{i_1}, \delta\chi_{i_2}, \ldots, \delta\chi_{i_k})
\ee
which are constant within some large box centered on the origin.

\subsection{Combining information from multiple sources}
\label{subsec:combining}
Although the detection rate for binary neutron stars is still rather uncertain, we expect advanced instruments to detect several events per year~\cite{ratespaper}. It is therefore important to take advantage of multiple detections to provide tighter constraints on the validity of GR.
Consider a set of $\mathcal{N}$ independent GW events, corresponding to $\mathcal{N}$ independent data sets $d_A$. We do not assume that deviations from GR are necessarily consistent between events, but rather that they can vary from source to source.
We assume there is a common underlying theory of gravity that describes emission of gravitational waves from the sources that are observed, but that shifts in the values of the parameters can vary from one source to another, over and above the dependence of the parameters on the masses.
One can write down a combined odds ratio for the catalog of
sources:
\ba
&&{}^{(N_T)}\mathcal{O}^{\rm modGR}_{\rm GR} \nn\\
&& = \frac{P(\hyp_{\rm modGR}|d_1, \dots, d_\mathcal{N},\info)}{P(\hyp_{\rm GR}|d_1, \dots, d_\mathcal{N},\info)} \nn\\
&& = \frac{\sum_{k=1}^{N_T} \sum_{i_1 < i_2 < \ldots < i_k} P(H_{i_1 i_2 \ldots i_k}|d_1, \dots, d_\mathcal{N},\info)}{P(\hyp_{\rm GR}|d_1, \dots, d_\mathcal{N},\info)} \nn\\
&& = \sum_{k=1}^{N_T} \sum_{i_1 < i_2 < \ldots < i_k} \frac{P(H_{i_1 i_2 \ldots i_k}|\info)}{P(\hyp_{\rm GR}|\info)}\,{}^{\rm (cat)}B^{i_1 i_2 \ldots i_k}_{\rm GR},
\label{step1}
\ea
where
\be
{}^{\rm (cat)}B^{i_1 i_2 \ldots i_k}_{\rm GR} = \frac{P(d_1, \dots, d_\mathcal{N}|H_{i_1 i_2 \ldots i_k},\info)}{P(d_1, \dots, d_\mathcal{N}|\hyp_{\rm GR},\info)}.
\ee
Since the events $d_1, \dots, d_\mathcal{N}$ are all independent, one has
\ba
P(d_1, \dots, d_\mathcal{N}|H_{i_1 i_2 \ldots i_k},\info) &=& \prod_{A=1}^{\mathcal{N}} P(d_A|H_{i_1 i_2 \ldots i_k},\info), \nn\\
P(d_1, \dots, d_\mathcal{N}|\hyp_{\rm GR},\info)          &=& \prod_{A=1}^{\mathcal{N}} P(d_A|\hyp_{\rm GR},\info).
\ea
Thus,
\be
{}^{\rm (cat)}B^{i_1 i_2 \ldots i_k}_{\rm GR} = \prod_{A=1}^{\mathcal{N}} {}^{(A)}B^{i_1 i_2 \ldots i_k}_{\rm GR},
\label{step2}
\ee
with
\be
{}^{(A)}B^{i_1 i_2 \ldots i_k}_{\rm GR} = \frac{P(d_A|H_{i_1 i_2 \ldots i_k},\info)}{P(d_A|\hyp_{\rm GR},\info)}.
\ee
To evaluate the combined odds ratio of the catalog we choose to invoke indifference (as in Eqs.~(\ref{restrictedprioroddsratios}) and (\ref{generalprioroddsratios})) and set the individual prior odds ratios equal to each other, so that
\be
\frac{P(H_{i_1 i_2 \ldots i_k}|\info)}{P(\hyp_{\rm GR}|\info)} = \frac{\alpha}{2^{N_T} - 1}.
\label{relativepriorodds}
\ee
Together with Eqns.~(\ref{step1}), (\ref{step2}), this leads to
\be
{}^{(N_T)}\mathcal{O}^{\rm modGR}_{\rm GR} =
\frac{\alpha}{2^{N_T} -1}\,\sum_{k=1}^{N_T} \sum_{i_1 < i_2 < \ldots < i_k} \prod_{A=1}^\mathcal{N} {}^{(A)}B^{i_1 i_2 \ldots i_k}_{\rm GR},
\label{oddscombined}
\ee
which, up to an overall prefactor, amounts to taking the average of the cumulative Bayes factors (\ref{step2}).

Alternatively, one may prefer not to make any assumptions for the prior odds ratios $P(H_{i_1 i_2 \ldots i_k}|\info)/P(\hyp_{\rm GR}|\info)$ at all, and
focus on the cumulative Bayes factors ${}^{\rm (cat)}B^{i_1 i_2 \ldots i_k}_{\rm GR}$ separately and individually. It will also be of interest to look at the cumulative Bayes factors for the various hypotheses $H_{i_1 i_2 \ldots i_k}$ and $\hyp_{\rm GR}$ against the noise-only hypothesis $\hyp_{\rm noise}$:
\ba
\prod_A {}^{(A)}B^{i_1 i_2 \ldots i_k}_{\rm noise} &=& \prod_A \frac{P(d_A|H_{i_1 i_2 \ldots i_k},\info)}{P(d_A|\hyp_{\rm noise},\info)}, \nn\\
\prod_A {}^{(A)}B^{\rm GR}_{\rm noise} &=& \prod_A \frac{P(d_A|\hyp_{\rm GR},\info)}{P(d_A|\hyp_{\rm noise},\info)},
\ea
and we note that
\be
{}^{\rm (cat)}B^{i_1 i_2 \ldots i_k}_{\rm GR} = \frac{\prod_A {}^{(A)}B^{i_1 i_2 \ldots i_k}_{\rm noise}}{\prod_{A'} {}^{(A')}B^{\rm GR}_{\rm noise}}.
\ee
Finally, we will also look at the individual contributions ${}^{(A)}B^{i_1 i_2 \ldots i_k}_{\rm noise}$ and ${}^{(A)}B^{\rm GR}_{\rm noise}$, for
$A = 1, 2, \ldots, \mathcal{N}$.

\section{Results}
\label{s:results}

To illustrate the method, we construct catalogs of 15 binary neutron star sources, distributed uniformly in
volume, with random sky positions and orientations, and whose total number is taken to be on the conservative side of the `realistic' estimates
in \cite{ratespaper} for the number of detectable sources in a one-year time span. We take the individual neutron star masses to lie between 1 and 2 $M_\odot$. The distance interval is between
100 Mpc and 400 Mpc; the former number is the radius within which one would expect $\sim0.5$ BNS inspirals per year,
and 400 Mpc is the approximate horizon distance in Advanced LIGO. The corresponding signals are added coherently
to stationary, Gaussian simulated data for the Advanced Virgo interferometer and the two Advanced LIGOs. A lower cut-off of 8 is implemented on the optimal network SNR, defined as the quadrature sum of individual detector SNRs. The analysis of the surviving signals
is performed with an appropriately modified version of the nested sampling code available in LAL \cite{LAL}.

For simplicity, we took the phase up to only 2PN order ($M = 4$), both in the signal and in the model waveforms. As an example, we
considered the case of $N_T = 3$ testing coefficients $\psi_1$, $\psi_2$, $\psi_3$.
Nested sampling then gives $2^3 - 1 = 7$ Bayes factors $B^1_{\rm GR}$, $B^2_{\rm GR}$, $B^3_{\rm GR}$, $B^{12}_{\rm GR}$, $B^{13}_{\rm GR}$, $B^{23}_{\rm GR}$, and $B^{123}_{\rm GR}$, which for a single source
can be combined to form the odds ratio (\ref{oddsindividual}). Note that up to and including 2PN order, the post-Newtonian coefficients take the general form \cite{mais10}
\be
\psi_i(\Mc,\eta) = \frac{3}{128 \eta} g_i(\eta)\,(\pi \Mc\,\eta^{-3/5})^{(i-5)/3},
\ee
where the $g_i(\eta)$ are polynomials in $\eta$; for the lowest three sub-leading PN orders one has
\be
g_1(\eta) \equiv 0,\,\,\,\,\,\,\,g_2(\eta) = \frac{20}{9}\left(\frac{743}{336} + \frac{11}{4}\eta\right),\,\,\,\,\,\,\,g_3(\eta) = -16 \pi.
\ee
Accordingly, in the model waveforms we allow for deformations
\ba
\psi_1^{\rm GR}(\Mc,\eta) = 0 &\rightarrow& \frac{3}{128\eta} (\pi \Mc\,\eta^{-3/5})^{-4/3} \delta\chi_1, \nn\\
\psi_2^{\rm GR}(\Mc,\eta) &\rightarrow&  \frac{3}{128\eta} g_2(\eta)\,(\pi \Mc\,\eta^{-3/5})^{-1}\left[1 + \delta\chi_2 \right],\nn\\
\psi_3^{\rm GR}(\Mc,\eta) &\rightarrow&  \frac{3}{128\eta} g_3(\eta)\,(\pi \Mc\,\eta^{-3/5})^{-2/3}\,\left[1 + \delta\chi_3 \right].\nn\\
\ea

We take the prior on $\delta\chi_i$ to be flat and centered on zero, with a total width of 0.5. This will be much larger than
the deviations we will use in simulated signals and hence suffices to illustrate the method; for real measurements one may want to
choose a still wider prior.

For the purposes of showing results, the factor $\alpha$ in Eqns.~(\ref{oddsindividual}) and (\ref{oddscombined}) will be set to one.

\subsection{Measurability of a deviation in a post-Newtonian coefficient}

In order to gauge the sensitivity of our method to deviations from GR, we first consider a large number of simulated signals with a constant relative offset $\delta\chi_3$. If GR is violated, we do not expect the deviation to be this simple, but these examples will serve to illustrate both the workings and the effectiveness of the technique for low-SNR sources of the kind we expect in Advanced LIGO and Virgo. We note that $\psi_3$ is the lowest-order coefficient which incorporates the non-linearity of General Relativity through so-called tail effects \cite{bs94,bs95}, and is therefore of particular interest.

\subsubsection{Signals with constant relative deviation $\delta\chi_3 = 0.1$}

We start with signals that have $\delta\chi_3 = 0.1$. We first compute the odds ratios for individual sources, ${}^{(N_T)}O^{\rm modGR}_{\rm GR}$, according to (\ref{oddsindividual}), with $N_T = 3$. Next we divide these up \emph{randomly} into catalogs of 15 sources each and compute the combined odds ratios ${}^{(N_T)}\mathcal{O}^{\rm modGR}_{\rm GR}$ as in (\ref{oddscombined}). We do the same thing
for injections that are pure GR, \emph{i.e.}~$\delta\chi_i = 0$ for all $i$, and again compute the quantities ${}^{(N_T)}O^{\rm modGR}_{\rm GR}$ for individual sources,
and ${}^{(N_T)}\mathcal{O}^{\rm modGR}_{\rm GR}$ for catalogs of 15 sources each.

\begin{figure}[h!]
\centering
\includegraphics[angle=0,width=\columnwidth]{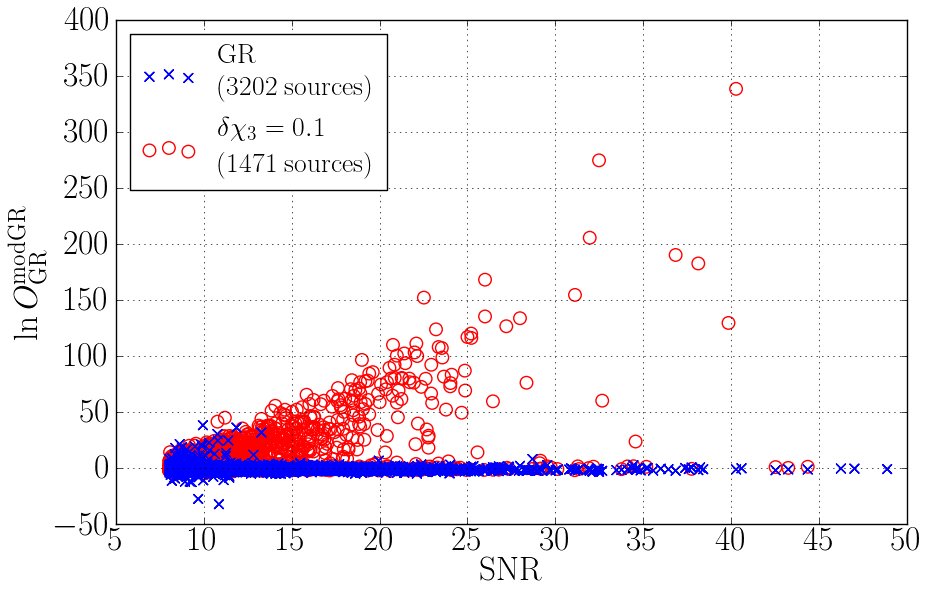}
\caption{The log odds ratios for individual sources. The blue crosses represent signals with standard GR waveforms, the red circles signals with a constant 10\% relative offset in $\psi_3$. A separation between the two is visible for SNR $\gtrsim 10$ and becomes more pronounced as the SNR increases.}
\label{fig:OvsSNR}
\end{figure}

Before considering catalogs, let us look at the (log) odds ratios for individual sources as a function of SNR. This is shown in Fig.~\ref{fig:OvsSNR}. The overwhelming majority of signals have SNR between 8 and 15, consistent with our SNR cut and the placement of sources uniformly in volume up to 400 Mpc. Even for low SNR sources, there is separation between the GR injections and the injections with modified $\psi_3$. As one would expect, the separation becomes much clearer with increasing SNR.

\begin{figure}[h!]
\centering
\includegraphics[angle=0,width=\columnwidth]{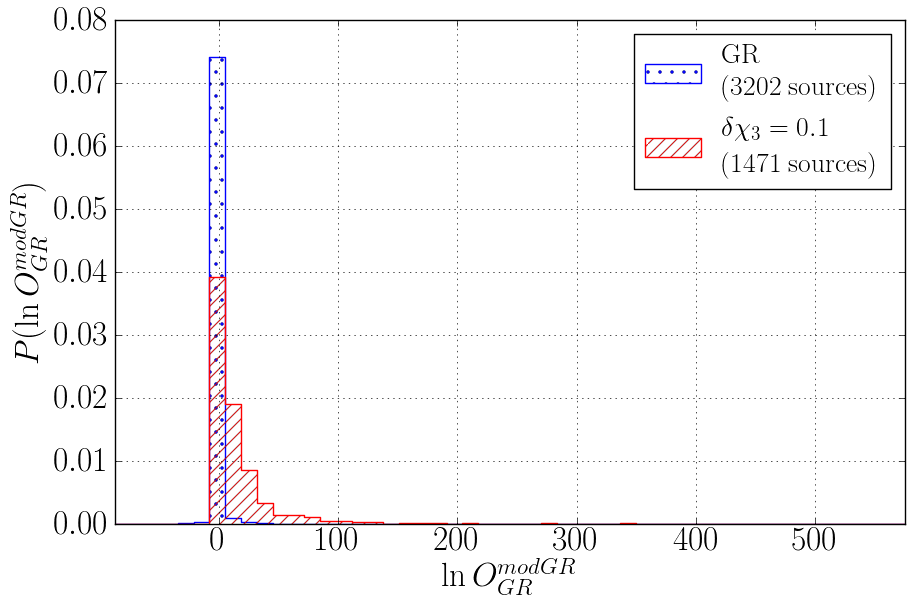}
\includegraphics[angle=0,width=\columnwidth]{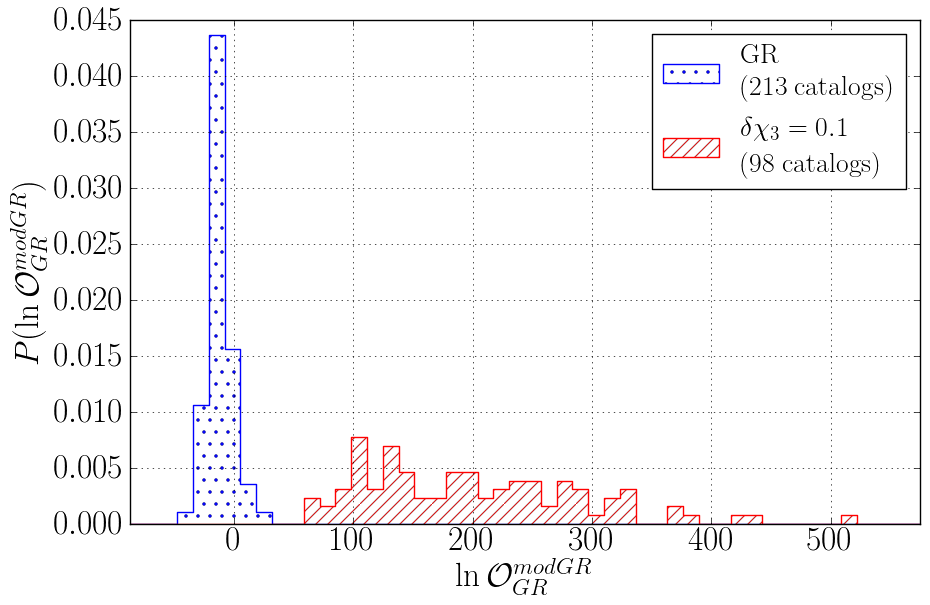}
\caption{Top: The normalized distribution $P(\ln O^{\rm modGR}_{\rm GR})$ of log odds ratios for individual sources, where the injections are either GR or have $\delta\chi_3 = 0.1$. Bottom: The normalized distribution $P(\ln \mathcal{O}^{\rm modGR}_{\rm GR})$ of logs of the combined odds ratios for GR injections and injections with $\delta\chi_3 = 0.1$, for catalogs of 15 sources each. The effectiveness of the catalog approach to testing for deviations from GR comes from the combination of multiple sources, each source contributing to the overall result in proportion to its own Bayes factors. 
}
\label{fig:dphi3_10pc}
\end{figure}

Fig.~\ref{fig:dphi3_10pc} shows normalized distributions of the log odds ratios, both for individual sources and for catalogs of 15 sources each. Combining the odds ratios for sources within a catalog will strongly boost our confidence in a violation of GR if one is present at the given level. For this particular choice of $\delta\chi_3$ in the injected non-GR signals, the separation from the GR injections is complete.

\begin{figure}[h!]
\centering
\includegraphics[angle=0,width=\columnwidth]{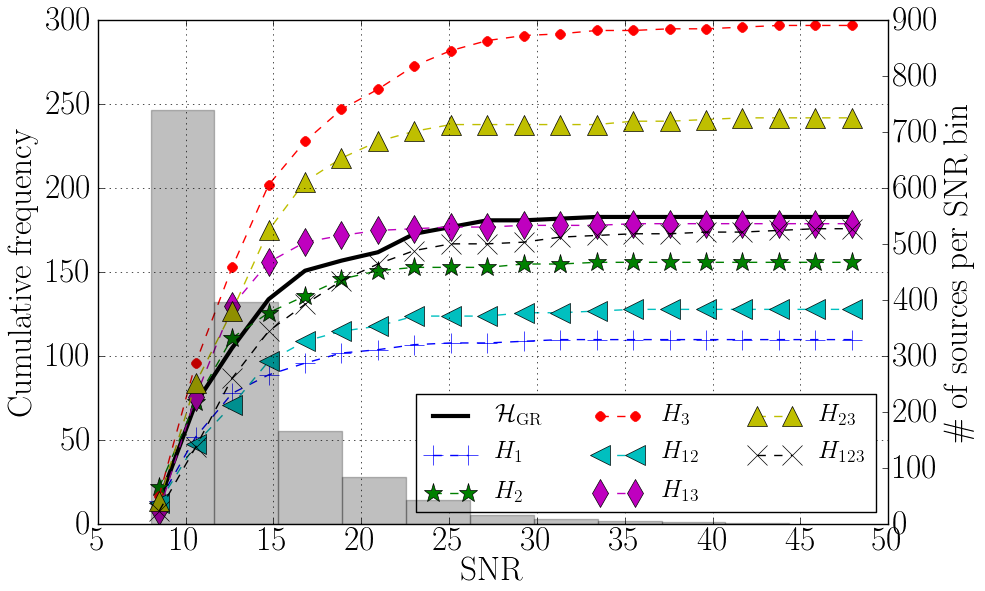}
\includegraphics[angle=0,width=\columnwidth]{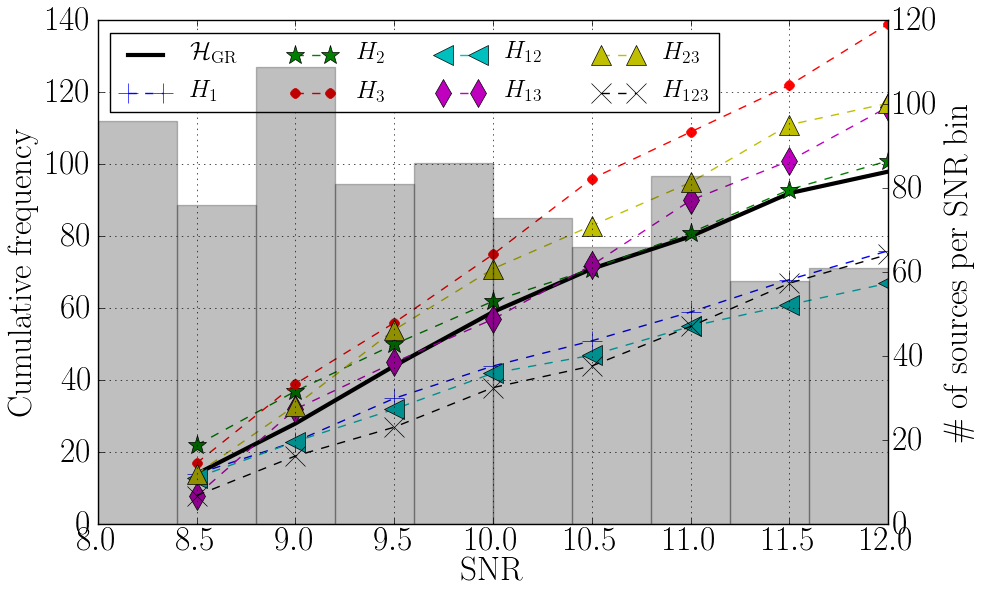}
\caption{
Top; curves and left vertical axis: For a given SNR, the cumulative number of times that the Bayes factor against noise for a particular component hypothesis is the largest for injections with that SNR or below, for $\delta\chi_3 = 0.1$. All 1471 simulated sources were used.
As expected, $B^3_{\rm noise}$ dominates, and Bayes factors for hypotheses that let $\psi_3$ be non-GR tend to outperform those that do not. The GR model outperforms the one with the largest number of free parameters. Histogram and right vertical axis: The number of sources per SNR bin.\\
Bottom: The same as above, but restricting to sources with SNR $< 12$. Similar behavior as for the full set of 1471 sources is observed. Note that already at SNR close to threshold, the GR hypothesis is more likely to be disfavored.
}
\label{fig:relativefrequencies}
\end{figure}

It is useful to look at which of the Bayes factors of the component hypotheses tend to give the largest contribution to the odds ratio. What is computed \emph{directly} by the nested sampling code is not $B^1_{\rm GR}, \ldots, B^3_{\rm GR}, B^{12}_{\rm GR}, \ldots, B^{23}_{\rm GR}, B^{123}_{\rm GR}$, but rather the Bayes factors for each of the hypotheses against the noise-only hypothesis $\hyp_{\rm noise}$:
\be
B^{i_1 \ldots i_k}_{\rm noise} = \frac{P(d|H_{i_1 \ldots i_k}, \info)}{P(d|\hyp_{\rm noise}, \info)},\,\,\,\,\,\,\,\,B^{\rm GR}_{\rm noise} = \frac{P(d|\hyp_{\rm GR}, \info)}{P(d|\hyp_{\rm noise}, \info)},
\ee
and one has
\be
B^{i_1 \ldots i_k}_{\rm GR} = \frac{B^{i_1 \ldots i_k}_{\rm noise}}{B^{\rm GR}_{\rm noise}}.
\ee
In Fig.~\ref{fig:relativefrequencies}, we show the cumulative number of times that a particular $B^{i_1 \ldots i_k}_{\rm noise}$ is the largest, against SNR, for the case where the injections have $\delta\chi_3 = 0.1$. The results are entirely as expected, considering that the injected waveform has a shift in $\psi_3$ only:
\begin{itemize}
\item The Bayes factor $B^3_{\rm noise}$ corresponding to the hypothesis $H_3$ dominates;
\item The Bayes factors $B^{i_1 \ldots i_k}_{\rm noise}$ corresponding to hypotheses that involve $\psi_3$ being non-GR tend to outperform those that do not;
\item The Bayes factors for the non-GR hypotheses deviate from the GR one already at low SNR, showing that our method will perform well in the low-SNR scenario.
\end{itemize}

Because of the first two points, one may be tempted to assign different prior odds to the various  hypotheses instead of setting them all equal to each other. For instance one might consider downweighting the most inclusive hypothesis, $H_{123}$, by invoking Occam's razor. However, the violation of GR we assume here is of a rather special form. In reality one will not know beforehand what the nature of the deviation will be; in particular, its effect may not be restricted to a single phase coefficient. It is possible that \emph{all} coefficients are affected, in which case one would not want to \emph{a priori} deprecate $H_{123}$. As explained in Sec.~\ref{subsec:general}, our hypothesis $\hyp_{\rm modGR}$ corresponds to the question whether \emph{one or more} of the phasing coefficients $\{\psi_1, \psi_2, \psi_3\}$ differ from their GR values; one may want to ask a different question, but this is the one that is the most general within our framework. To retain full generality, all sub-hypotheses $H_{i_1 i_2 \ldots i_k}$ need to be taken into account and given equal weight.



\subsubsection{Signals with constant relative deviation $\delta\chi_3 = 0.025$}

It is clear that, if signals arriving at the Advanced Virgo-LIGO network would have a (constant) fractional deviation in $\psi_3$ as large as 10\%, then
at least under the assumption of Gaussian noise, we would have no trouble in discerning this violation of GR even if only 15 events were ever recorded. Now let us look at a smaller deviation in $\psi_3$; say, 2.5\%.

\begin{figure}[h!]
\centering
\includegraphics[angle=0,width=\columnwidth]{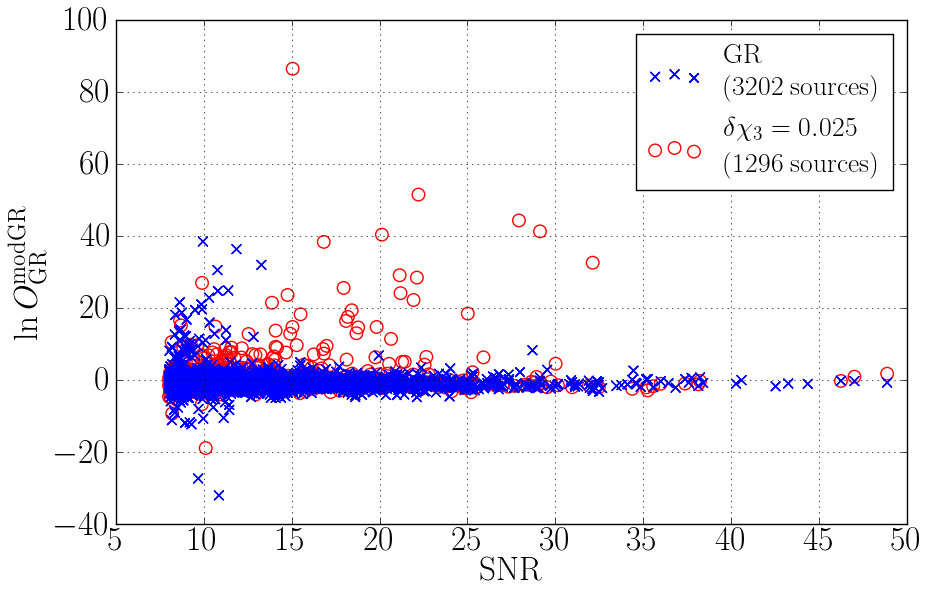}
\caption{The log odds ratios for individual sources. The blue crosses are for signals with standard GR waveforms, the red circles for signals with a constant 2.5\% relative offset in $\psi_3$. This time there is little separation between the two, although there are outliers for SNR $\gtrsim 15$.}
\label{fig:OvsSNR_2p5pc}
\end{figure}

In Fig.~\ref{fig:OvsSNR_2p5pc} we plot the log odds ratios for individual sources against SNR, both for signals with GR waveforms and signals with $\delta\chi_3 = 0.025$. This time the two distributions largely coincide, although there are some outliers which could boost the \emph{combined} odds ratio
when they are present in a catalog of sources.

\begin{figure}[h!]
\centering
\includegraphics[angle=0,width=\columnwidth]{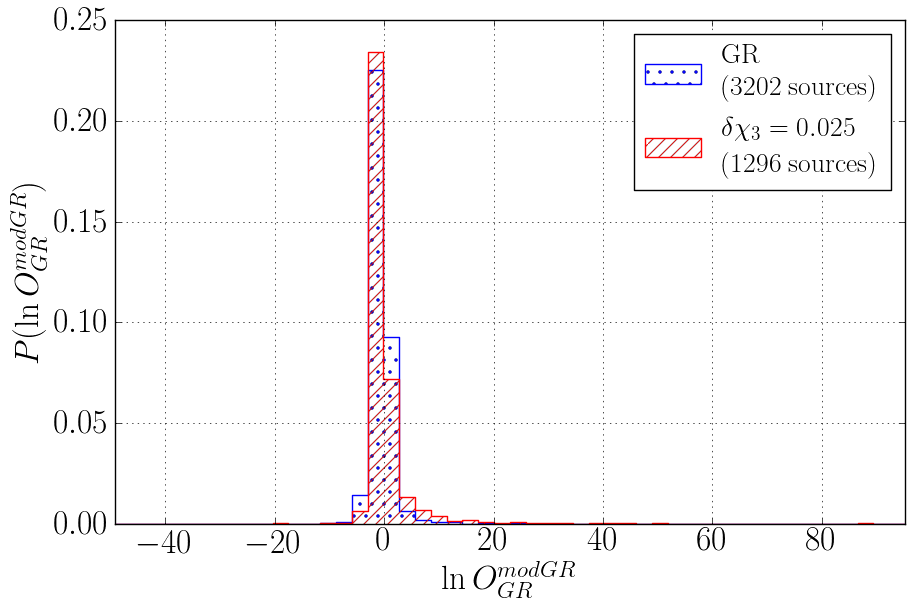}
\includegraphics[angle=0,width=\columnwidth]{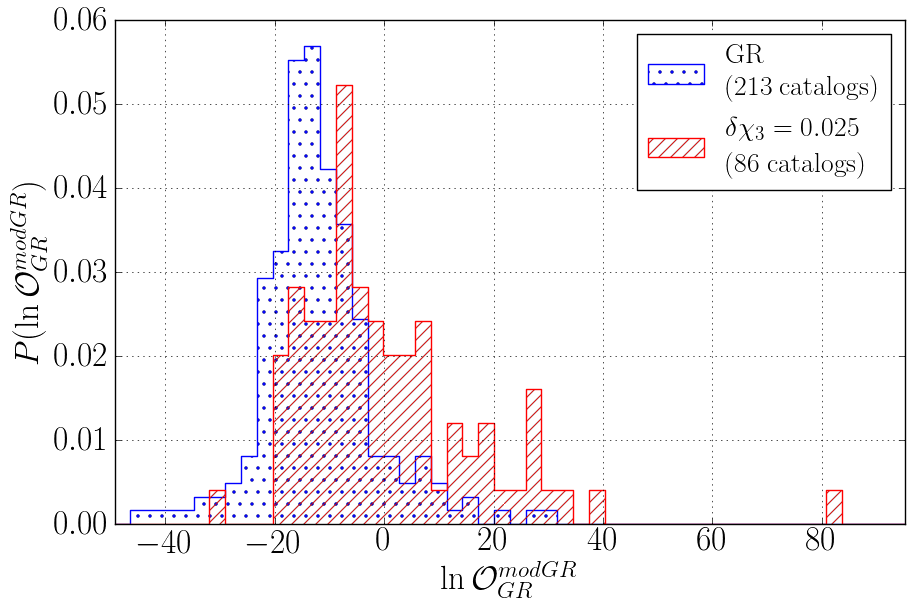}
\caption{Top: The normalized distribution $P(\ln O^{\rm modGR}_{\rm GR})$ of log odds ratios for individual sources, where the injections are either GR or have $\delta\chi_3 = 0.025$. Bottom: The normalized distribution $P(\ln \mathcal{O}^{\rm modGR}_{\rm GR})$ of logs of the combined odds ratios for GR injections and injections with $\delta\chi_3 = 0.025$, for catalogs of 15 sources each. For individual sources, the two distributions essentially lie on top of each other. However, when sources are combined into catalogs, it is possible for an outlier to boost the odds ratio of the entire catalog.}
\label{fig:dphi3_2p5pc}
\end{figure}

In Fig.~\ref{fig:dphi3_2p5pc} we show normalized distributions of the log odds ratio for individual sources, as well as for catalogs with 15 sources each. For individual sources, the distributions are more or less on top of each other. The picture is somewhat different for the catalogs. If a catalog with $\delta\chi_3 = 0.025$ happens to contain one of the outliers visible in Fig.~\ref{fig:OvsSNR_2p5pc}, then it can boost the combined odds ratio for the catalog.

\begin{figure}[h!]
\centering
\includegraphics[angle=0,width=\columnwidth]{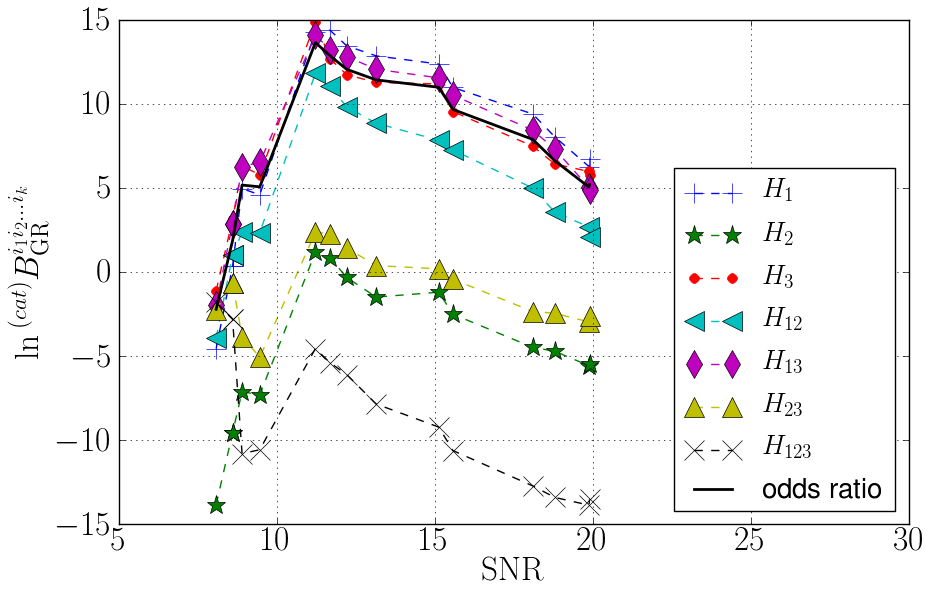}
\caption{The build-up of cumulative Bayes factors against GR for individual hypotheses, and the odds ratio, for a typical catalog with $\delta\chi_3 = 0.025$. Note that on the basis of the Bayes factor against GR of the most inclusive hypothesis $H_{123}$ alone, one would have concluded that the GR model is in fact the favored one. Even the log Bayes factor for $H_{23}$ ends up being negative. Additionally, the hypothesis with the largest Bayes factor is not $H_3$ but $H_1$. This illustrates that it is necessary to include as many hypotheses as possible in the analysis.}
\label{fig:cat_2p5pc}
\end{figure}

It is instructive to look at a representative catalog with $\ln {}^{(3)}\mathcal{O}^{\rm modGR}_{\rm GR} > 0$. In Fig.~\ref{fig:cat_2p5pc} we show the build-up of the log Bayes factors for the various sub-hypotheses against GR, as well as the odds ratio itself. In a scenario where the evidence for a GR violation is marginal, it is imperative to include as many hypotheses as possible in the analysis. Indeed, in this example, the hypothesis $H_3$ is not the most favored one; instead, it is $H_1$. Note also that if we had only tested the most inclusive hypothesis $H_{123}$ against GR (as one might do if one expects a deviation in all of the PN parameters), we would have concluded that the GR hypothesis is the favored one. The same is true if we had only tested $H_2$, as one would do when specifically looking for a `massive graviton'. Even the log Bayes factor for $H_{23}$ ends up being negative. We also remind the reader that we will not know beforehand what the precise nature of the GR violation will be.

By using the distribution of log odds ratios for simulated catalogs of GR sources, one can establish a \emph{threshold} which the odds ratio of a given catalog must overcome in order that a violation of GR becomes credible. This would be the analog of what was done in \cite{vv09} (see Fig.~7 of that paper), where the distribution of the log Bayes factor $\ln B_{\rm S,N}$ for the presence of a signal versus noise-only was computed for many realizations of the noise, in the absence of a signal. Consider the distribution $P\left(\ln {}^{(N_T)}\mathcal{O}^{\rm modGR}_{\rm GR}|\kappa, \hyp_{\rm GR}, \info\right)$ of log odds ratio for the collection $\kappa$ of simulated catalogs of signals that are in accordance with GR. Given a `false alarm probability' $\beta$, a threshold $\ln \mathcal{O}_\beta$ for the odds ratio can be set as follows:
\be
\beta = \int_{\ln \mathcal{O}_\beta}^\infty P(\ln \mathcal{O}|\kappa,\hyp_{\rm GR}, \info)\,d \ln\mathcal{O}.
\ee
Now suppose we also have a distribution $P\left(\ln {}^{(N_T)}\mathcal{O}^{\rm modGR}_{\rm GR}|\kappa', \hyp_{\rm alt}, \info\right)$ of log odds ratio for a collection $\kappa'$ of simulated catalogs of signals which follow some alternative theory (in this example, one which leads to a shift $\delta\chi_3 = 0.025$). Then we can quantify the chance that a deviation from GR of this particular kind will be detected with a false alarm probability smaller than the given $\beta$, by means of an \emph{efficiency} $\zeta$, defined as
\be
\label{eq:efficiency}
\zeta = \int_{\ln\mathcal{O}_\beta}^\infty P(\ln\mathcal{O}|\kappa', \hyp_{\rm alt}, \info)\,d\ln\mathcal{O}.
\ee
We note that with these definitions, the efficiency is independent of the overall prior odds of $\hyp_{\rm modGR}$ versus $\hyp_{\rm GR}$, as the factor $\alpha$ in Eqns.~(\ref{totalignorance}) and (\ref{oddscombined}) will just cause both the distributions $P\left(\ln {}^{(N_T)}\mathcal{O}^{\rm modGR}_{\rm GR}|\kappa, \hyp_{\rm GR}, \info\right)$ and $P\left(\ln {}^{(N_T)}\mathcal{O}^{\rm modGR}_{\rm GR}|\kappa', \hyp_{\rm alt}, \info\right)$, and the threshold $\ln \mathcal{O}_\beta$, to be shifted by $\ln \alpha$.

In the present example, with $\delta\chi_3 = 0.025$ and catalogs of 15 sources each, for $\beta = 0.05$ one has $\zeta = 0.22$. Hence, by this standard and for the given number of sources in a catalog, a GR violation of this kind is just borderline detectable.

\begin{figure}[h!]
\centering
\includegraphics[angle=0,width=\columnwidth]{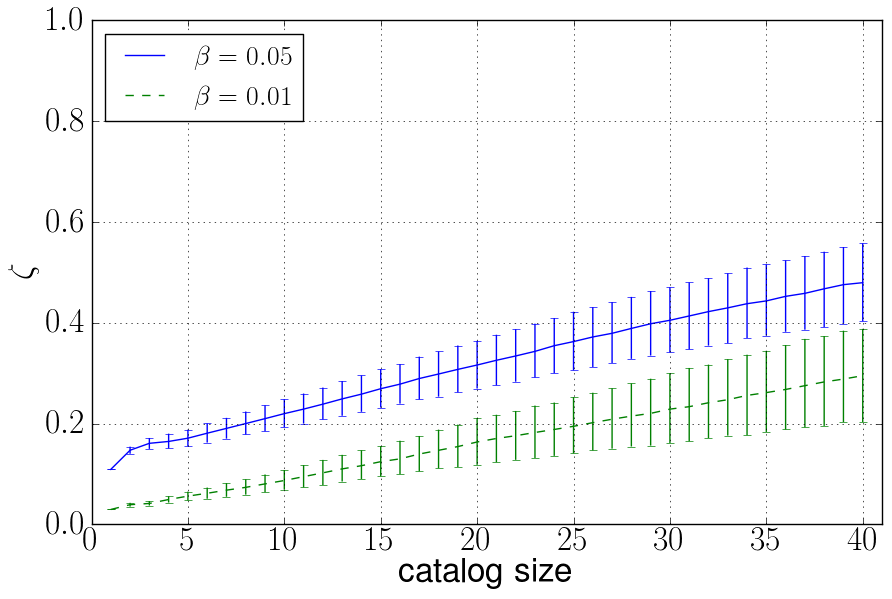}
\caption{The efficiency of detecting a GR violation for sources with $\delta\chi_3 = 0.025$, as a function of catalog size for false alarm probabilities $\beta \in \{0.05, 0.01\}$. The median and the $68\%$ confidence interval from 5000 random catalog orderings are shown as the central curve and the error bars, respectively. The efficiency increases as a function of catalog size, once again underscoring the benefit of combining all available data.}
\label{fig:efficiency_dphi3_2p5pc}
\end{figure}

However, the number of sources per catalog, 15 in the examples shown so far, and the chosen false alarm probability, $\beta = 0.05$, are somewhat arbitrary. In reality the size of the catalog will depend on the number of detected sources, and the false alarm probability is set according to the required confidence. It is therefore of interest to investigate the effects of both of these factors. In Fig.~\ref{fig:efficiency_dphi3_2p5pc} we show the behavior of the efficiency as a function of the catalog size and the false alarm probability. To account for the arbitrariness in which the sources are combined to form a catalog, we show the median and $68\%$ confidence interval of the efficiencies from 5000 random source orderings as the central curves and the error bars, respectively. Results are shown for $\beta \in \{0.05,0.01\}$.

As is evident from Fig.~\ref{fig:efficiency_dphi3_2p5pc}, the efficiency rises as a function of the catalog size. This highlights the importance of combining all available sources in the advanced detector era, when one looks for deviations from GR. The maximum catalog sizes shown are comparable to the `realistic' estimates of the number of detections of binary neutron star inspirals in a the span of a year \cite{ratespaper}.

We see that $\delta\chi_3 = 0.025$ is a borderline case in terms of discernability of a GR violation. Later on, when we show posterior PDFs, it will become evident that indeed, $\delta\chi_3$ can typically be measured with an an accuracy of this order.

\subsection{Effect of number of hypotheses used}

It is of interest to see what would have happened if we had used a smaller number of testing coefficients; say, $\{\psi_1,\psi_2\}$, so that the hypotheses to be tested are $H_1$, $H_2$, and $H_{12}$. In the example with $\delta\chi_3 = 0.025$ as in the previous subsection, the PN order where the deviation occurs, namely 1.5PN, would then be higher than the PN orders associated with our testing coefficients, which are 0.5PN and 1PN. 

\begin{figure}[h!]
\centering
\includegraphics[angle=0,width=\columnwidth]{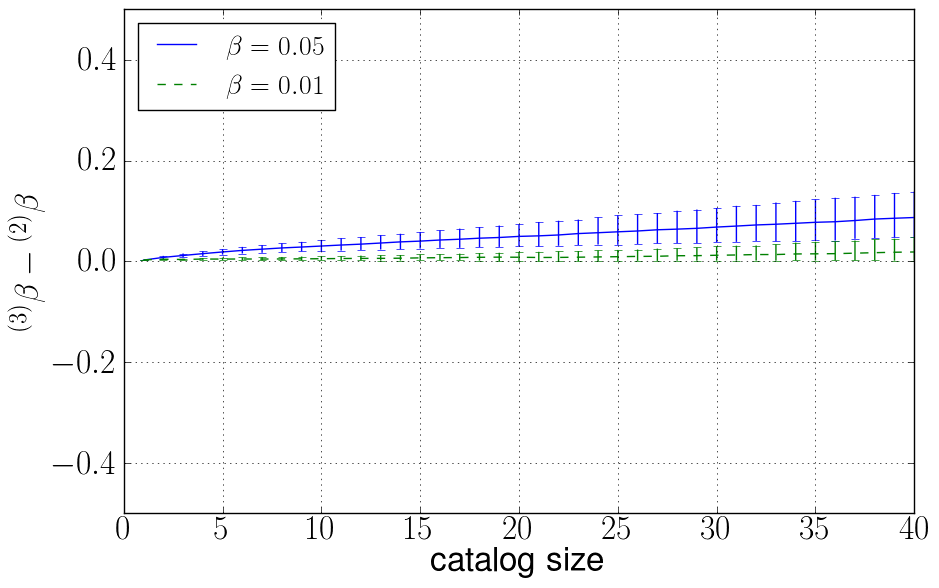}
\includegraphics[angle=0,width=\columnwidth]{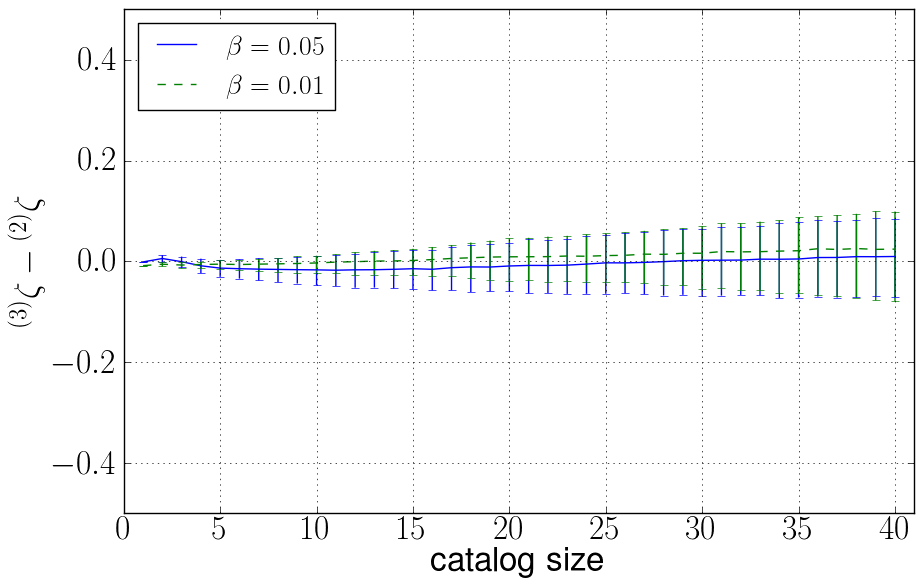}
\caption{Top: The change in false alarm probabilities (FAPs) in going from two to three testing parameters, but keeping the odds ratio thresholds fixed. Bottom: The change in efficiencies when keeping the false alarm probabilities fixed. The plots shown are for the case where the signals have $\delta\chi_3 = 0.025$. We see that increasing the number of testing parameters has only a moderate effect on the FAPs, while for fixed FAPs, the efficiencies do no change appreciably. Note, however, that when the evidence for a deviation from GR is marginal, the use of as many hypotheses as possible can be pivotal in finding the violation (see Fig.~\ref{fig:cat_2p5pc}, and also Fig.~\ref{fig:catalogs}).}
\label{fig:FAP_and_Efficiency}
\end{figure}

In Fig.~\ref{fig:FAP_and_Efficiency}, the following two things are shown:
\begin{itemize}
\item In the case where only $\{\psi_1,\psi_2\}$ are testing coefficients, we compute the thresholds $\ln {}^{(2)}\mathcal{O}_\beta$ corresponding to false alarm probabilities (FAPs) $\beta \in \{0.05, 0.01\}$. Next, we re-calculate the false alarm probabilities \emph{for the same thresholds}, but now for the case where there are three testing parameters, $\{\psi_1, \psi_2, \psi_3\}$, and show the difference in false alarm probabilities;
\item On the other hand, one can compare the efficiencies ${}^{(2)}\zeta$ and ${}^{(3)}\zeta$ for the two and three parameter cases, \emph{for fixed false alarm probabilities} $\beta \in \{0.05, 0.01\}$. This is shown in the bottom panel.  
\end{itemize}

As expected, in the first case (fixed thresholds for the odds ratios), the false alarm probabilities increase in going from two to three testing parameters, but only moderately so. On the other hand, for fixed false alarm probabilities, there is no appreciable change in efficiency. Indeed, the spread in the GR `background' will increase with an increase in hypotheses to be tested against GR; yet, having more hypotheses does not really hurt us in terms of our ability to detect a deviation from GR. 

Fig.~\ref{fig:FAP_and_Efficiency} indicates the typical behavior for catalogs with a \emph{specific} deviation from GR, in this case $\delta\chi_3 = 0.025$. 
It is worth repeating, however, that especially when there is only marginal evidence for a GR violation, it is important to use as many hypotheses as is computationally feasible; see Fig.~\ref{fig:cat_2p5pc} (and also Fig.~\ref{fig:catalogs} below). Also, we will obviously not know beforehand what the \emph{nature} of the GR violation is.

\begin{figure}[h!]
\centering
\includegraphics[angle=0,width=\columnwidth]{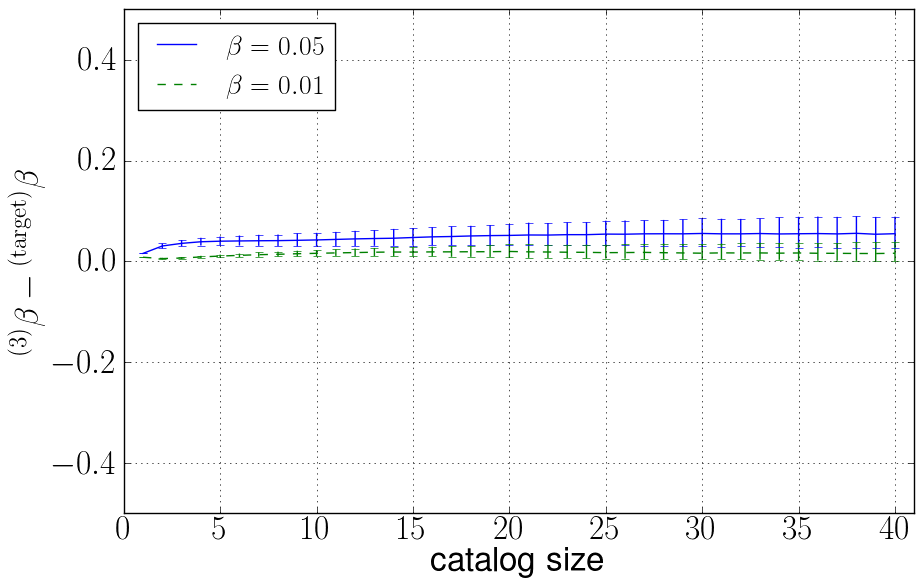}
\includegraphics[angle=0,width=\columnwidth]{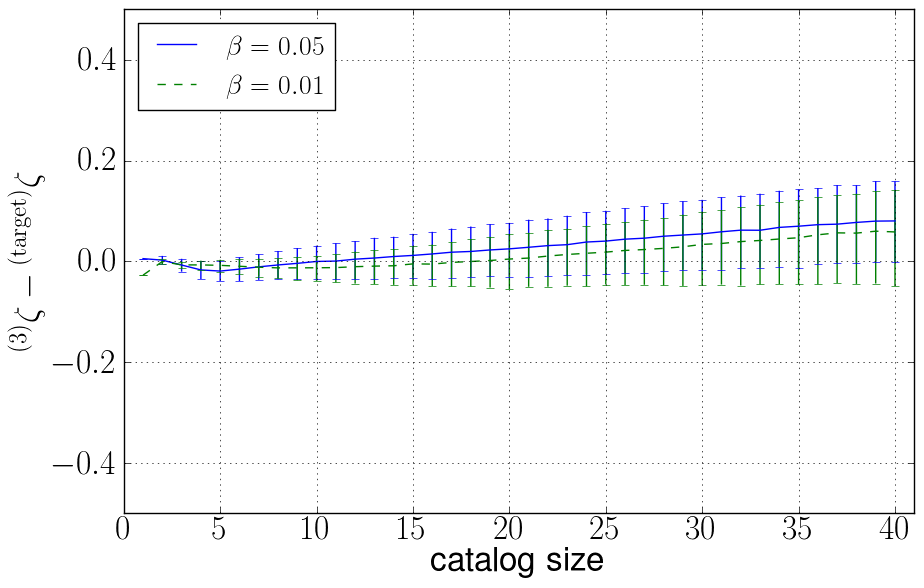}
\caption{Top: The difference of false alarm probabilities, ${}^{(3)}\beta - {}^{({\rm target})}\beta$, for fixed log odds ratio thresholds and signals having $\delta\chi_3 = 0.025$, between our 3-parameter test and a `targeted search' which only looks for a deviation in $\psi_3$, \emph{i.e.}, only tests the hypothesis $H_3$ against GR. We see that the difference
is minor.
Bottom: More important is the difference in efficiencies, ${}^{(3)}\zeta - {}^{({\rm target})}\zeta$, for fixed false alarm probabilities. Especially for a large number of sources per catalog, our 3-parameter test is actually more efficient than the `targeted search', at least for this particular example. This is because the Bayes factor against GR for $H_3$ will not be the largest in every catalog, but our method naturally compensates for that. Of course, we do not expect our method to outperform a targeted search in the case of a more complicated deviation from GR.}
\label{fig:ourtest_vs_H3}
\end{figure}

One may nevertheless wonder how our 3-parameter test would compare with a `targeted search' that only looks for a deviation in $\psi_3$, which in this example happens to be where the deviation actually is. With our choice of $\alpha = 1$, this corresponds to setting $\mathcal{O}^{\rm modGR}_{\rm GR} = {}^{\rm (cat)}B^3_{\rm GR}$. Fig.~\ref{fig:ourtest_vs_H3} shows the change in false alarm probabilities in going from testing only $H_3$ to the full test for fixed log odds ratio thresholds, as well as the change in efficiencies for fixed false alarm probabilities. The results are as follows:
\begin{itemize}
\item The change in false alarm probabilities for fixed log odds ratio thresholds is minor;
\item However, especially for a large number of sources per catalog, the efficiencies show a clear rise. This can be accounted for by the fact that, for a small violation of GR, it will not always be the case that the Bayes factor against GR for $H_3$ is the largest, but our method is able to compensate for that.
\end{itemize}
We conclude that for this \emph{particular} example, our method with $N_T = 3$ testing parameters will tend to outperform a `targeted search' that happens to look for the violation actually present. However, we do not expect this to be true for more complicated deviations from GR.

\subsection{Measurability of deviations with non-PN frequency dependences}

The aim of the previous subsections was to get a rough idea of the sensitivity of our method to deviations in post-Newtonian coefficients, and in order to
gauge this we assumed a constant relative offset in the physically interesting parameter $\psi_3$. However, we stress once again that we do not expect a violation of GR to manifest itself as a simple constant relative shift in one of the post-Newtonian coefficients.
Even if modifications are confined to the PN coefficients, the $\delta\chi_i$ in the signals can be dependent on $(\Mc,\eta)$, in addition to whatever charges
and coupling constants may be present. Moreover, a deviation from GR could introduce terms in the phase with frequency dependences that do not correspond to
any of the PN contributions. We now show that the method can also be sensitive to violations of that kind, even though the model waveforms we use in our analyses only have deformations of PN terms.
Let us give an heuristic example where the phase of the simulated signals contains a term with an anomalous frequency dependence in between that of the 1PN and 1.5PN contributions. Specifically,
\be
\frac{3}{128\eta}(\pi \Mc \eta^{-3/5})^{-5/6} \delta\chi_A f^{-5/6},
\label{anomalous}
\ee
and we note that the 1PN term goes like $f^{-1}$ and the 1.5PN term like $f^{-2/3}$; thus, the deviation introduced here could be dubbed `1.25PN'. However, for the recovery, we will continue to use the same model waveforms as before, which can only have shifts in the phase coefficients at 0.5PN, 1PN, and 1.5PN. Our aim in this subsection is to show that they will nevertheless allow us to find a deviation in the signal of the form (\ref{anomalous}).

We now need to make a choice for $\delta\chi_A$. We aim to show that even if there is a deviation in the phase that is not represented in any of our model waveforms, it can be recovered, if near the `bucket' of the noise curve (at $f \sim 150$ Hz) the amount by which it affects the phase is on a par with a shift in the PN coefficients of more than a few percent.

For definiteness, let us take $\delta\chi_A$ to be constant, and such that at $f = 150$ Hz and for a system of $(1.5,1.5)\,M_\odot$, the contribution (\ref{anomalous}) to the phase is equal to the change caused by a shift in the 1.5PN contribution with $\delta\chi_3 = 0.1$:
\ba
&&(\pi 3\,M_\odot)^{-5/6} \delta\chi_A (150\,\mbox{Hz})^{-5/6} \nn\\
&&= g_3(0.25)\,(\pi 3\,M_\odot)^{-2/3} \times 0.1 \times (150\,\mbox{Hz})^{-2/3},
\ea
leading to $\delta\chi_A = -2.2$.

\begin{figure}[h!]
\centering
\includegraphics[angle=0,width=\columnwidth]{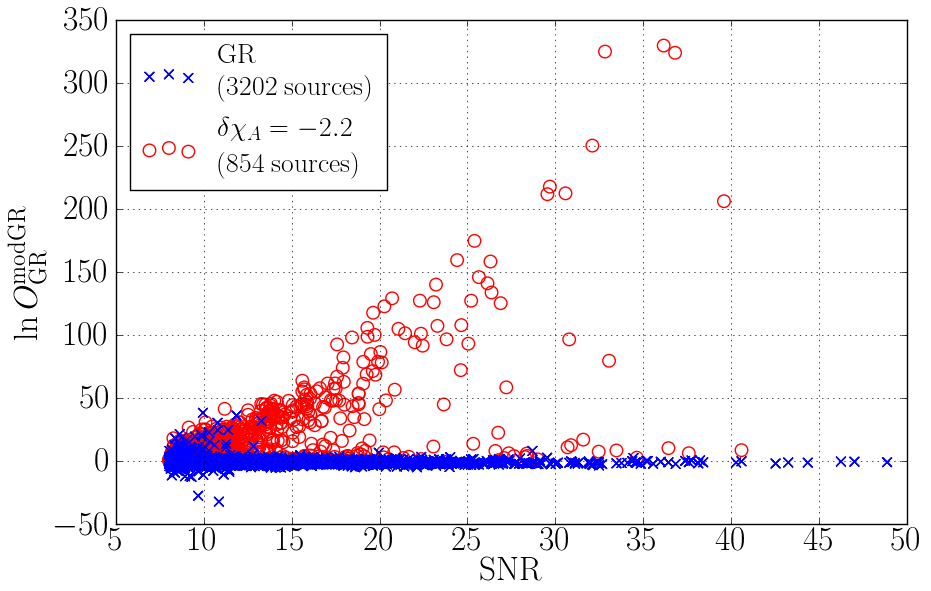}
\caption{The log odds ratios for individual sources for injections with an anomalous frequency dependence in the phase. The blue crosses represent GR injections, while the red circles are for signals that have a contribution to the phase with a frequency dependence in between that of the 1PN and 1.5PN terms (`1.25PN'). As with the $\delta\chi_3 = 0.1$ example, a separation between the two is visible for SNR $\gtrsim 10$ and becomes more pronounced as the SNR increases.}
\label{fig:OvsSNR_dphiA}
\end{figure}

As before, we first give results for the odds ratios of individual sources with increasing SNR; see Fig.~\ref{fig:OvsSNR_dphiA}. We see that even at small SNR there is already a good separation between the GR injections and the injections with a modification in the structure of the phase.

\begin{figure}[h!]
\centering
\includegraphics[angle=0,width=\columnwidth]{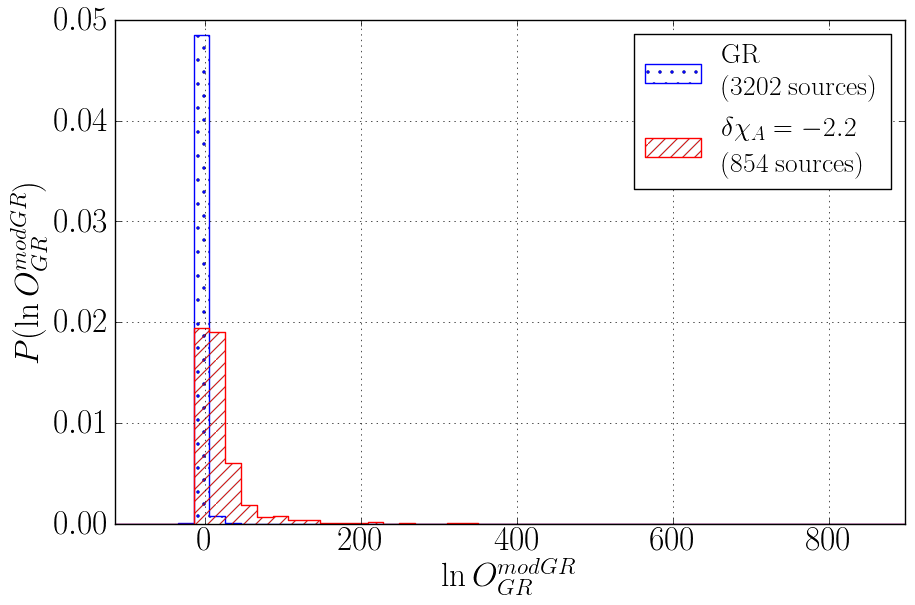}
\includegraphics[angle=0,width=\columnwidth]{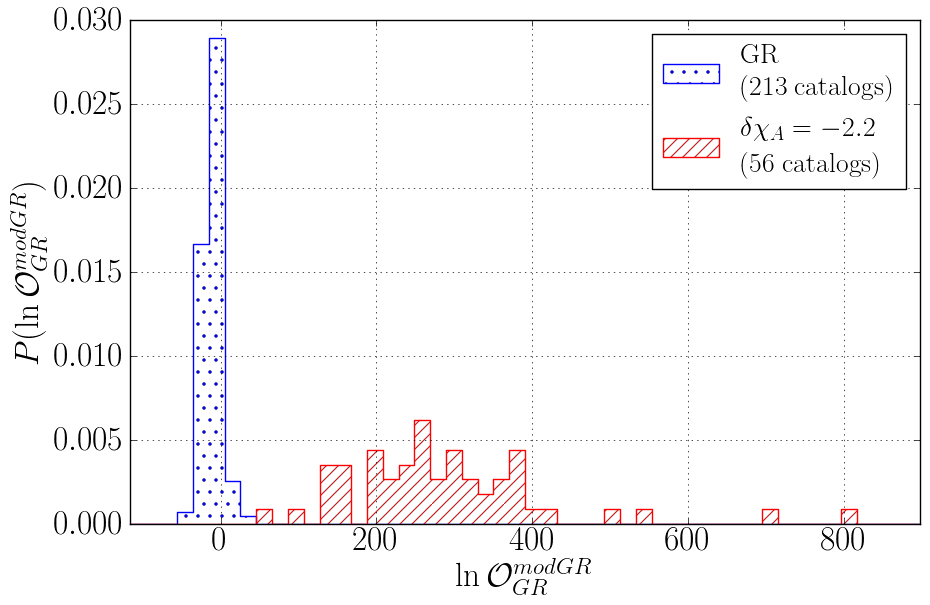}
\caption{Top: The normalized distribution $P(\ln O^{\rm modGR}_{\rm GR})$ of log odds ratios for individual sources, where the injections are either GR or have the anomalous frequency dependence. Bottom: The normalized distribution $P(\ln \mathcal{O}^{\rm modGR}_{\rm GR})$ of logs of the combined odds ratios for GR injections and injections for catalogs of 15 sources each.}
\label{fig:dphiA}
\end{figure}

Next we show normalized distributions of the log odds ratios, both for individual sources and for catalogs of 15 sources each: Fig.~\ref{fig:dphiA}. As expected, for the catalogs there is an excellent separation between the GR injections and injections with a modification in the phase.

\begin{figure}[h!]
\centering
\includegraphics[angle=0,width=\columnwidth]{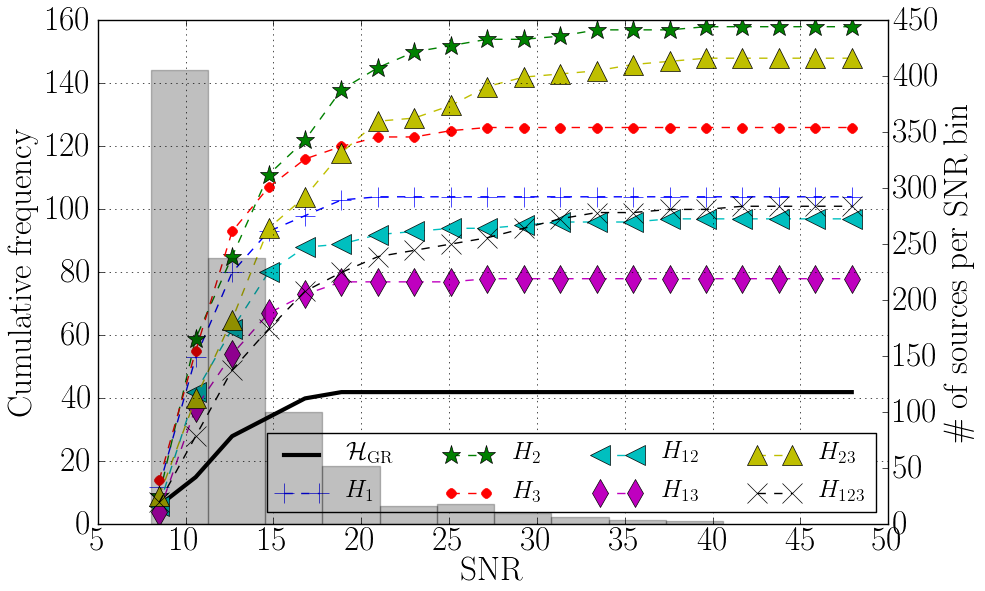}
\includegraphics[angle=0,width=\columnwidth]{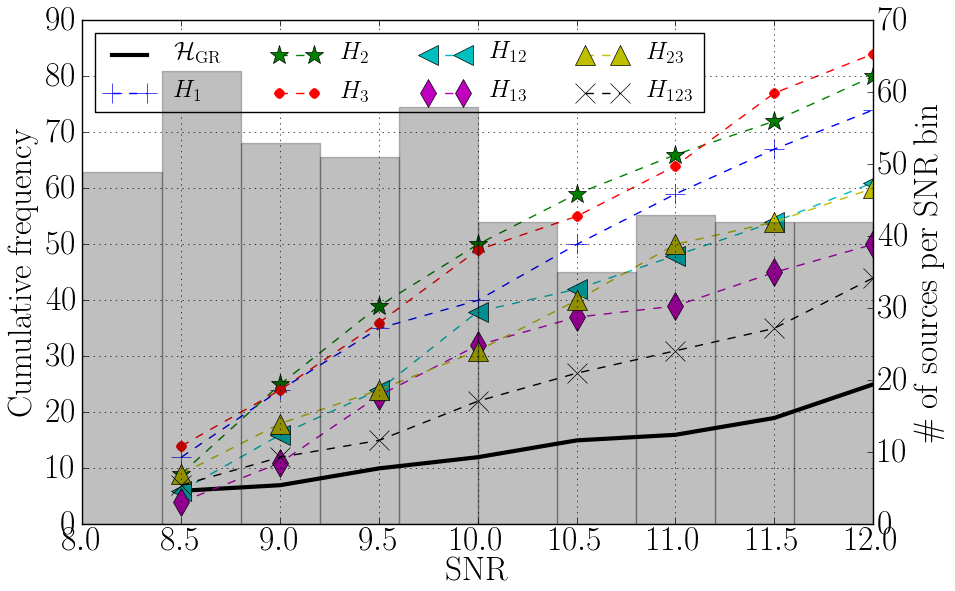}
\caption{Top; curves and left vertical axis: The cumulative number of times that the Bayes factor against noise for a particular component hypothesis is the largest, with increasing SNR, for individual sources with $\delta\chi_A = -2.2$. All 854 simulated sources were used. Histogram and right vertical axis: The number of sources per SNR bin.\\
Bottom: The same as above, but now for sources with SNR $< 12$. Close to threshold, no single hypothesis is the dominant one. 
}
\label{fig:relativefrequencies_dphiA}
\end{figure}

Now let us look at the cumulative number of times that the Bayes factor against noise for a particular hypothesis is the largest, for individual sources with $\delta\chi_A = -2.2$,  arranged with increasing SNR (Fig.~\ref{fig:relativefrequencies_dphiA}). From SNR $\gtrsim 15$, the Bayes factor $B^2_{\rm noise}$ starts to dominate, followed by $B^{23}_{\rm noise}$ and $B^3_{\rm noise}$, with the latter two crossing over between SNR $\sim 20$ and SNR $\sim 25$. Already at SNR $\sim 9$, \emph{all} of the $B^{i_1 i_2 \ldots i_k}_{\rm noise}$ dominate the Bayes factor $B^{\rm GR}_{\rm noise}$ for the GR hypothesis. However, near the SNR threshold, no single hypothesis dominates clearly, which again shows that as many hypotheses as possible should be included in the analysis.

\begin{figure}[h!]
\centering
\includegraphics[angle=0,width=\columnwidth]{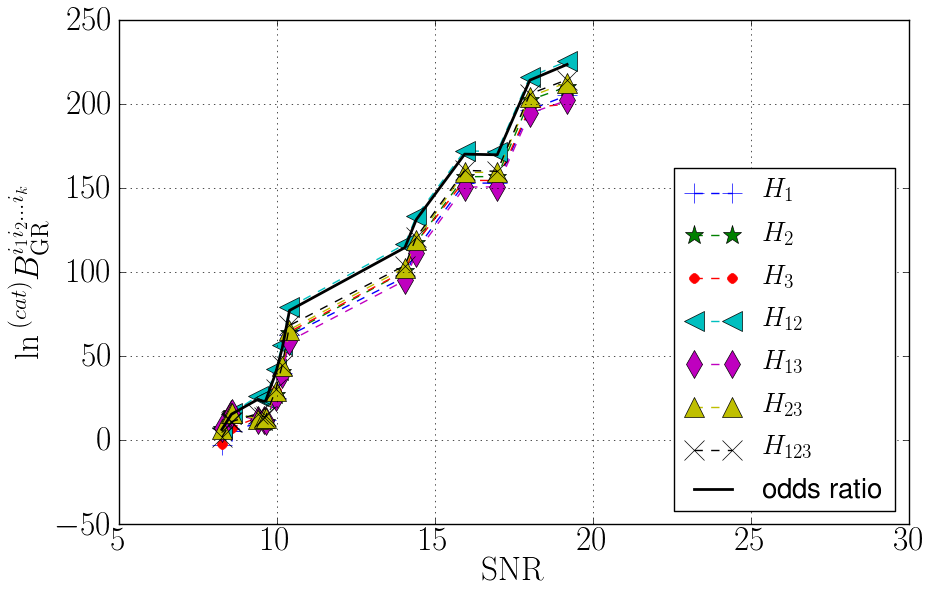}
\includegraphics[angle=0,width=\columnwidth]{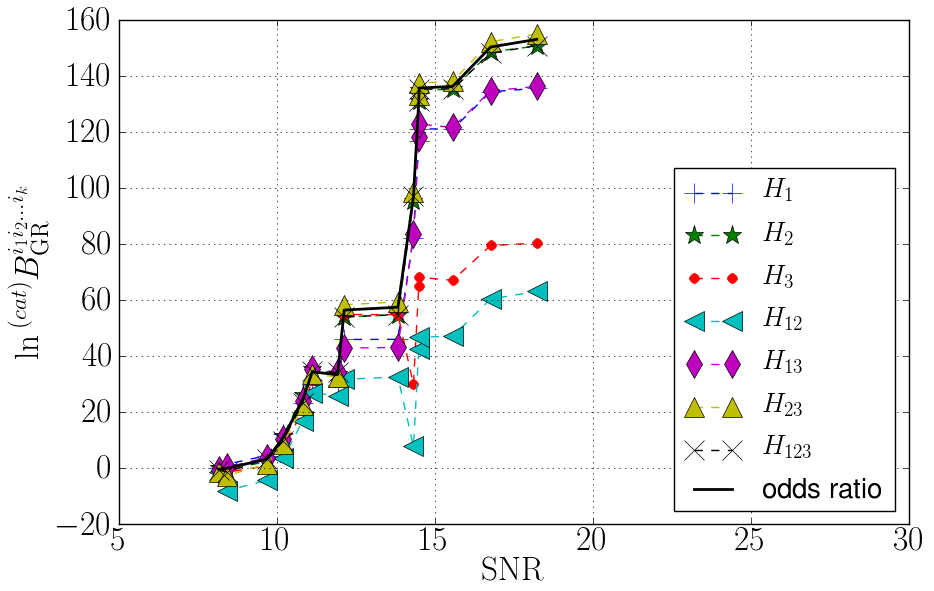}
\includegraphics[angle=0,width=\columnwidth]{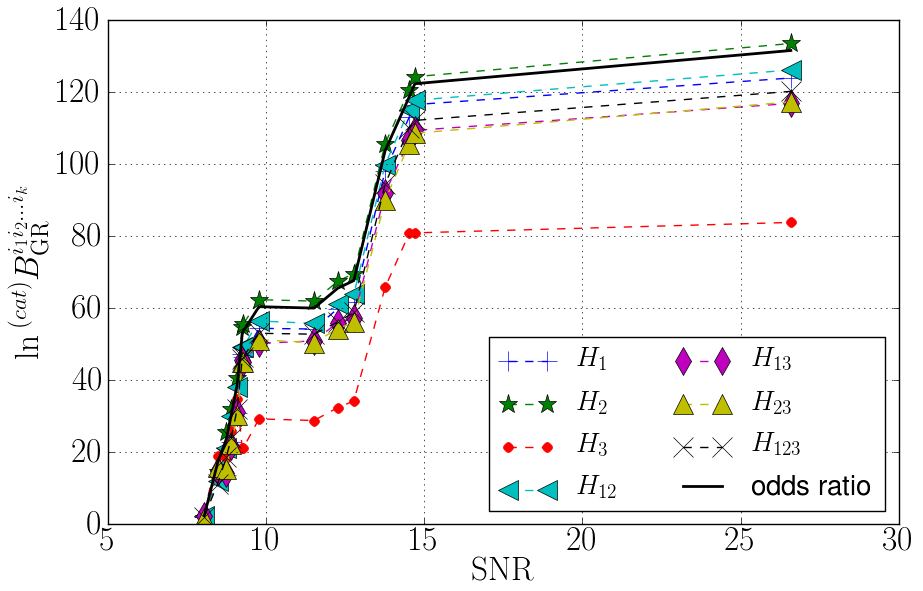}
\caption{A few examples of how cumulative Bayes factors against GR for individual hypotheses, and the odds ratio, grow as sources with increasing SNR are being added within three different catalogs of 15 sources in total. Top and middle: Catalogs where sources have only modest SNRs ($< 20$). Note the large differences in contributions from different hypotheses, and in the ordering of Bayes factors, between these two catalogs. Bottom: A catalog with a high SNR source; note however that the source with the highest SNR does not cause a particularly large `boost', and the cumulative log odds ratio ends up being considerably lower than in the top and middle examples.
}
\label{fig:catalogs}
\end{figure}

Especially in this case, it is interesting to look at the growth of cumulative Bayes factors against GR for individual hypotheses, as well as of the odds ratio, as sources with increasing SNR are being added within catalogs of 15 sources. This is shown for a few example catalogs in Fig.~\ref{fig:catalogs}.
The salient features are:
\begin{itemize}
\item Even if all 15 sources only have modest SNR, by their cumulative contributions they can cause a relatively large odds ratio for the catalog as a whole;
\item In catalogs containing a source with a particularly hight SNR, it is by no means a given that the contribution of this source will dominate the odds ratio compared to the cumulative contributions of the other sources;
\item Which hypothesis comes out on top will vary from one catalog to another; in the examples of Fig.~\ref{fig:catalogs} we see $H_{12}$, $H_2$, or $H_{23}$ giving the largest contribution, respectively, but there are examples where any of the other four sub-hypotheses contributes the most.
In this respect we note that the odds ratio for a catalog is proportional to the average of the cumulative Bayes factors themselves, not of their logarithms. If one were to \emph{a priori} favor particular (subsets of) hypotheses, the log odds ratio could be lowered by as much as 100. This could have a large effect on the false alarm probability; see Fig.~\ref{fig:dphiA}. These are again arguments for using as many sub-hypotheses as possible, and give them equal relative prior odds.
\end{itemize}

Note that in principle we could have extended our model waveforms with more free parameters so as to be more sensitive to deviations of this kind, \emph{e.g.}~by including a term in the phase of the form $\mathcal{A} f^a$, similar to what one has in the so-called parameterized post-Einsteinian (ppE) waveforms \cite{yp09}. However, the point of our method is to search for \emph{generic} deviations from GR. In the future, we will want to search with more sophisticated (time domain) waveforms whose use is computationally more demanding in a Bayesian setting, and we will not be able to allow for an arbitrarily large number of deformations in the model waveforms.
However, here we have an example where the injections have a contribution to the phase that is not present in any of the model waveforms, yet is clearly observable. Recall that the overall magnitude of the anomalous phase contribution was chosen so that, at $f \sim 150$ Hz, it changes the phase by a similar amount as a shift in $\psi_3$ of 10\%. Our results make it plausible that generically, when there is a deviation in the phase which, at frequencies where the detectors are the most sensitive, causes a phase change on a par with a change in one of the (low order) PN coefficients of more than a few percent, it will be detectable.

\subsection{Parameter estimation}

Finally, let us look at some posterior PDFs for the $\delta\chi_i$. We stress that unlike
Bayes factors, the PDFs can not be combined across sources since we should not expect the
$\delta\chi_i$ to be independent of the component masses; they can differ from source to source.
Even looking at the PDFs for a single source may then be misleading: even if the deviation is
exactly in one or more of the PN coefficients, a given source will have values for the $\delta\chi_i$ that are
representative just for the $(\Mc,\eta)$ of that source, and possibly also the values of additional charges that may
appear in an alternative theory of gravity. In a given catalog, there may be only one source
with sufficient SNR to allow for accurate parameter estimation, in which case the posteriors will not tell us
much even if they are strongly peaked. More generally, the deviation from GR may manifest itself
by the appearance of terms in the phase that do not have the frequency dependence of any of the PN contributions.
However, in the event that the log odds ratio and Bayes factors strongly favor the GR hypothesis, posterior PDFs
will allow us to constrain deviations in the PN coefficients, thereby adding further support that GR is
the correct theory. Hence we start with an analysis of pure GR injections.

\subsubsection{Example: A GR injection}

Let us first look at a GR source with $(\Mc,\eta,D) = (1.31\,M_\odot, 0.243, 131\,\mbox{Mpc})$, and a LIGO-Virgo network SNR of 23.0.
The Bayes factors for the various component hypotheses are:
\ba
\ln B^1_{\rm GR} = -2,\,\,\,\,\,\ln B^2_{\rm GR} &=& -2,\,\,\,\,\,\ln B^3_{\rm GR} = -2, \nn\\
\ln B^{12}_{\rm GR} =-3,\,\,\,\,\,\ln B^{13}_{\rm GR} &=& -1,\,\,\,\,\,\ln B^{23}_{\rm GR} = -1, \nn\\
\ln B^{123}_{\rm GR} &=& -2. \nn\\
\ea
We recall that these are the Bayes factors for a particular deviation from GR versus GR. The GR hypothesis
is favored in all cases. We can also look at the Bayes factors for all of the hypotheses
against noise:
\ba
\ln B^{\rm GR}_{\rm noise} &=& 211,\nn\\
\ln B^1_{\rm noise} = 209,\,\,\,\,\,\ln B^2_{\rm noise} &=& 209,\,\,\,\,\,\ln B^3_{\rm noise} = 209,\nn\\
\ln B^{12}_{\rm noise} = 208,\,\,\,\,\,\ln B^{13}_{\rm noise} &=& 210,\,\,\,\,\,\ln B^{23}_{\rm noise} = 210, \nn\\
\ln B^{123}_{\rm noise} &=& 209. \nn\\
\ea
Hence the signal is picked up very well by the waveforms of all of the hypotheses, with the GR waveform doing slightly better.

\begin{figure}[h!]
\centering
\includegraphics[angle=0,width=\columnwidth]{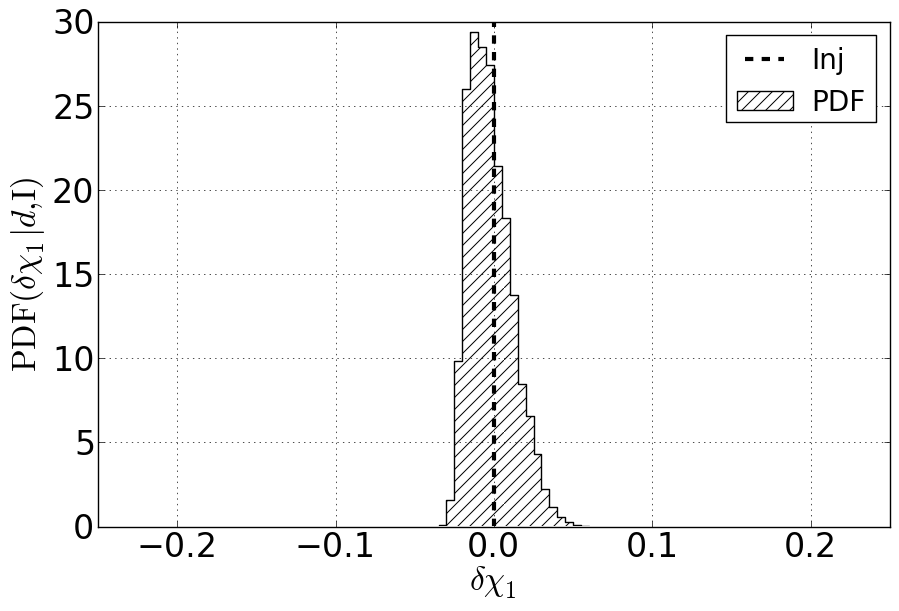}
\includegraphics[angle=0,width=\columnwidth]{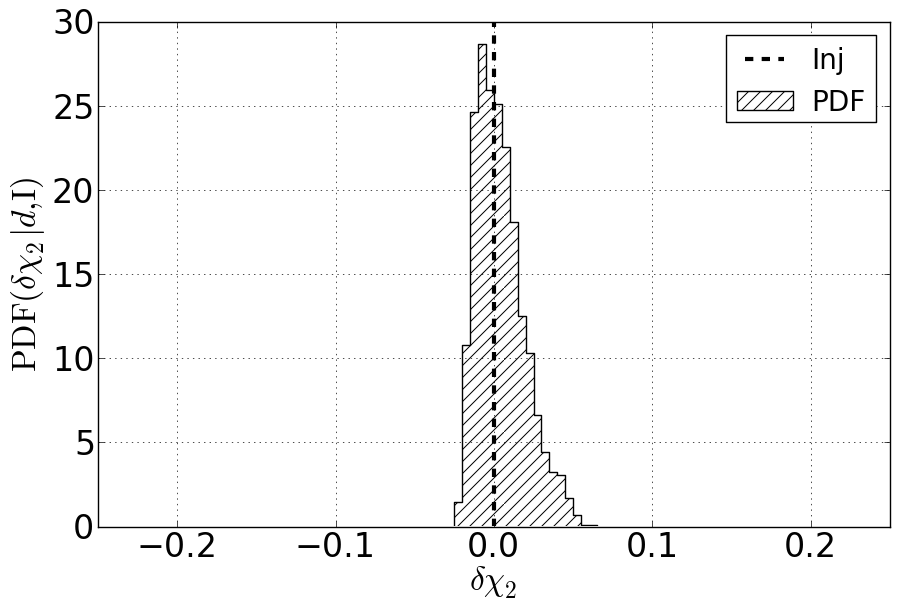}
\includegraphics[angle=0,width=\columnwidth]{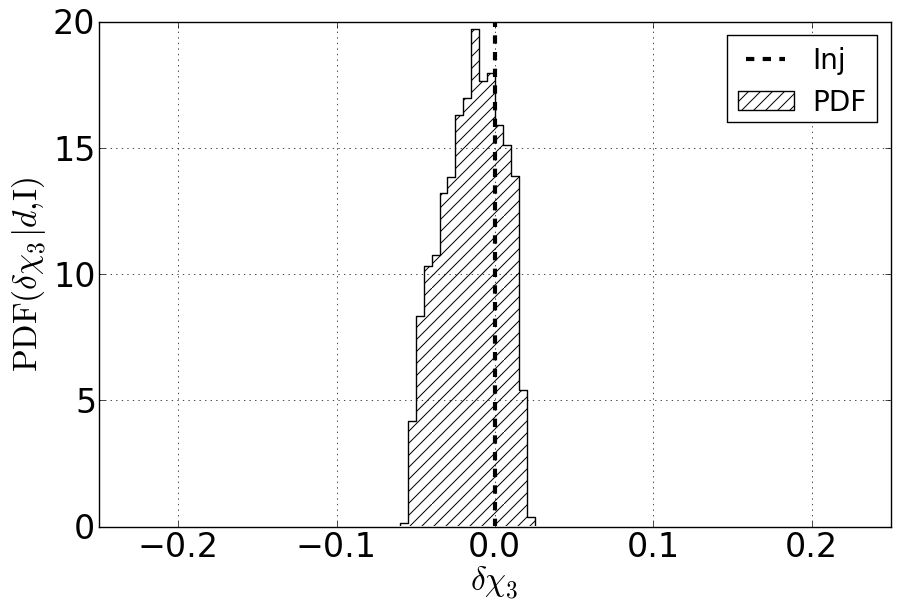}
\caption{Posterior PDFs for a single GR injection with network SNR of 23.0. Top: $\delta\chi_1$ measured with a waveform that has $\{\vec{\theta},\delta\chi_1\}$
as free parameters; middle: $\delta\chi_2$ measured with a waveform that has $\{\vec{\theta},\delta\chi_2\}$ free;
bottom: $\delta\chi_3$ measured with a waveform that has $\{\vec{\theta},\delta\chi_3\}$ free. In each case
the distribution is tightly centered on zero, with standard deviations of 0.014, 0.015, and 0.019, respectively.}
\label{fig:GR3hyps}
\end{figure}

Let us now look at some posterior PDFs. In Fig.~\ref{fig:GR3hyps}, we show the PDFs for $\delta\chi_1$, $\delta\chi_2$, and $\delta\chi_3$,
respectively for the waveforms that have free parameters $\{\vec{\theta}, \delta\chi_1\}$, $\{\vec{\theta},\delta\chi_2\}$, and $\{\vec{\theta},\delta\chi_3\}$,
with $\vec{\theta}$ the parameters of the GR waveform. We see that the distributions are all narrowly peaked around the correct value of zero.

\subsubsection{Example: A signal with $\delta\chi_3 = 0.1$.}

We now consider an example with $(\Mc,\eta,D) = (1.18\,M_\odot, 0.244, 196\,\mbox{Mpc})$, with a non-zero relative shift in $\psi_3$ of
 $\delta\chi_3 = 0.1$, and network SNR 23.2. The
Bayes factors are:
\ba
\ln B^1_{\rm GR} = 117,\,\,\,\,\,\ln B^2_{\rm GR} &=& 124,\,\,\,\,\,\ln B^3_{\rm GR} = 124, \nn\\
\ln B^{12}_{\rm GR} = 123,\,\,\,\,\,\ln B^{13}_{\rm GR} &=& 124,\,\,\,\,\,\ln B^{23}_{\rm GR} = 125, \nn\\
\ln B^{123}_{\rm GR} &=& 114.
\ea
This time the GR hypothesis is very much disfavored. However, we note that the Bayes factor
for the hypothesis that only $\psi_3$ differs from its GR value is not the largest. In fact, all
the Bayes factors except for $B^1_{\rm GR}$ and $B^{123}_{\rm GR}$ are rather similar in magnitude, and
no clear conclusions can be drawn from them regarding the underlying nature of the deviation from GR.

When looking at the Bayes factors against noise, we see that the signal is clearly detected for all hypotheses:
\ba
\ln B^{\rm GR}_{\rm noise} &=& 128,\nn\\
\ln B^1_{\rm noise} = 245,\,\,\,\,\,\ln B^2_{\rm noise} &=& 252,\,\,\,\,\,\ln B^3_{\rm noise} = 252,\nn\\
\ln B^{12}_{\rm noise} = 251,\,\,\,\,\,\ln B^{13}_{\rm noise} &=& 252,\,\,\,\,\,\ln B^{23}_{\rm noise} = 253, \nn\\
\ln B^{123}_{\rm noise} &=& 242. \nn\\
\ea

\begin{figure}[h!]
\centering
\includegraphics[angle=0,width=\columnwidth]{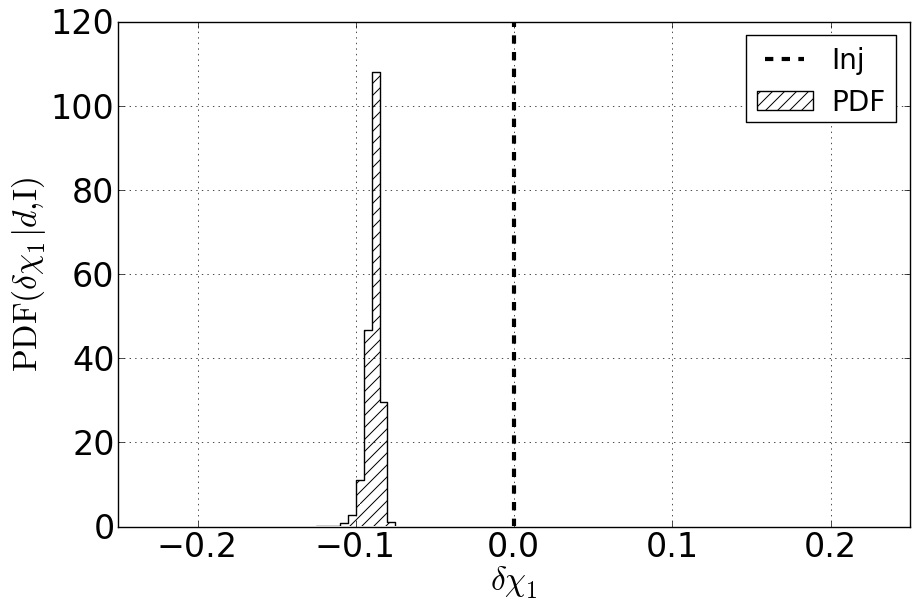}
\includegraphics[angle=0,width=\columnwidth]{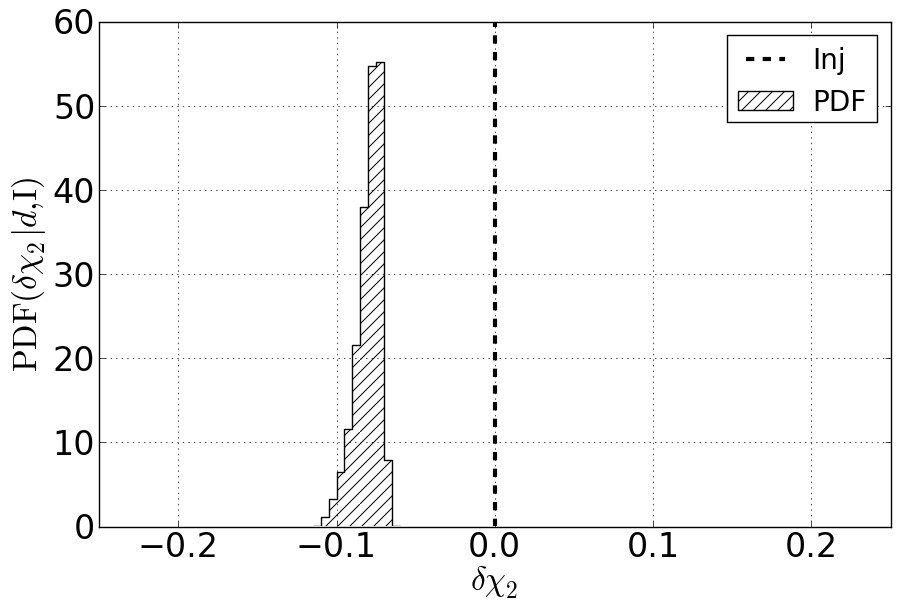}
\includegraphics[angle=0,width=\columnwidth]{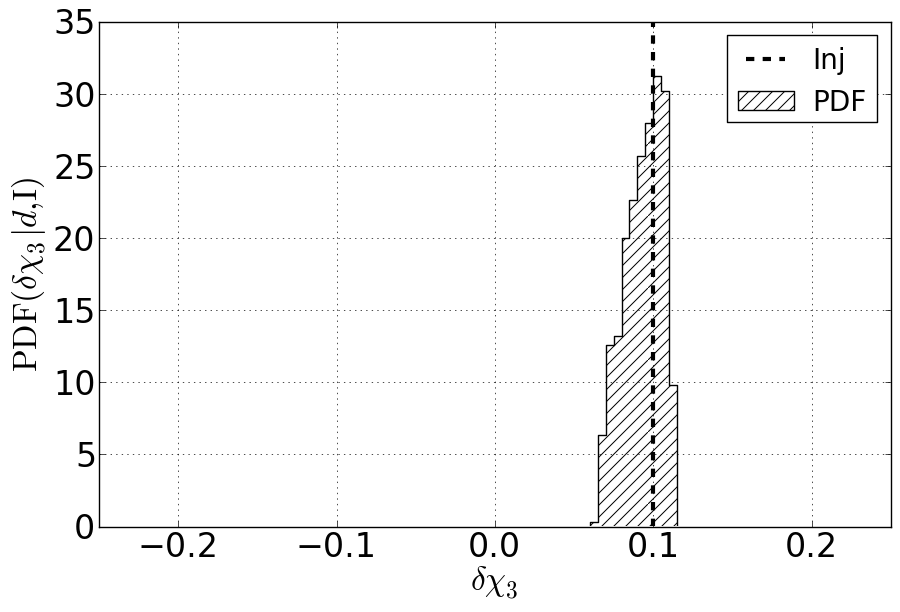}
\caption{Posterior PDFs for a single injection with $\delta\chi_3 = 0.1$ and network SNR 23.2. Top: $\delta\chi_1$ measured with a waveform that has $\{\vec{\theta},\delta\chi_1\}$ free; middle: $\delta\chi_2$ measured with a waveform that has $\{\vec{\theta},\delta\chi_2\}$ free; bottom: $\delta\chi_3$ measured with a waveform that has $\{\vec{\theta},\delta\chi_3\}$ free. As expected, the bottom PDF is sharply peaked at the correct value of $\delta\chi_3 = 0.1$, with a standard deviation of 0.012. For the top and middle PDFs, the one test parameter that is used differs from the parameter in the signal that has the shift. The parameters in the waveform will rearrange themselves such as to best accommodate the properties of the signal. Both $\delta\chi_1$ and $\delta\chi_2$ end up being sharply peaked, but not at the correct value of zero.}
\label{fig:dpsi3dpsi3}
\end{figure}

Now let us consider posterior PDFs. We expect
the PDF of $\delta\chi_3$ for the hypothesis $H_3$, where
only $\{\vec{\theta},\delta\chi_3\}$ are allowed to vary, to be peaked at
the injected value of 0.1, and this is the case with very good accuracy, as shown in the bottom panel of Fig.~\ref{fig:dpsi3dpsi3}.

In the upper and middle panels of the same figure, the PDFs of $\delta\chi_1$ for the hypothesis $H_1$, and of $\delta\chi_2$ for
the hypothesis $H_2$ are shown.
In these cases the parameter in the signal that has the shift is now not represented; in the first case only $\delta\chi_1$ is allowed to vary on top of the
parameters $\vec{\theta}$ of GR, and in the second case only $\delta\chi_2$. In the nested sampling process, the waveform will still try to adapt itself to the deformation in the
signal. The result is that $\delta\chi_1$ and $\delta\chi_2$ are strongly peaked, but away from the correct values $\delta\chi_1 = \delta\chi_2 = 0$. Thus, if one were to study the data only using waveforms from a \emph{specific} alternative theory of gravity (\emph{e.g.} a `massive graviton' model with a deviation in $\psi_2$ only), one might find a violation of GR but draw the wrong conclusions about the nature of the deviation.

\begin{figure}[h!]
\centering
\includegraphics[angle=0,width=\columnwidth]{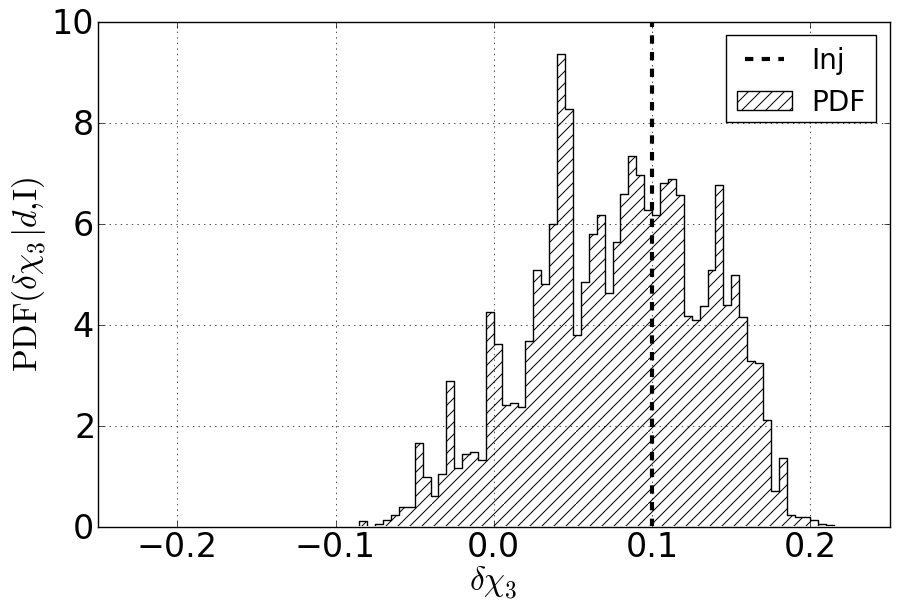}
\caption{The posterior PDF for the injection with $\delta\chi_3 = 0.1$ as in the previous figure, but recovered with waveforms where
$\{\vec{\theta},\delta\chi_1,\delta\chi_2,\delta\chi_3\}$ are all free. The peak is near the correct value of $\delta\chi_3$ (with a median of 0.083),
but this time the spread is considerably larger (with a standard deviation of 0.055), as we are trying to measure more parameters at the same time.}
\label{fig:dpsi3_allfree}
\end{figure}

We can also look at the PDF for the hypothesis $H_{123}$, where the waveforms have
$\delta\chi_1$, $\delta\chi_2$, $\delta\chi_3$ free; see Fig.~\ref{fig:dpsi3_allfree}. Once again the peak
is more or less at the correct value of $\delta\chi_3$, but we now have a much bigger spread. This too is as expected;
parameter estimation degrades if one tries to measure too many parameters at once.

\begin{figure}[h!]
\centering
\includegraphics[angle=0,width=\columnwidth]{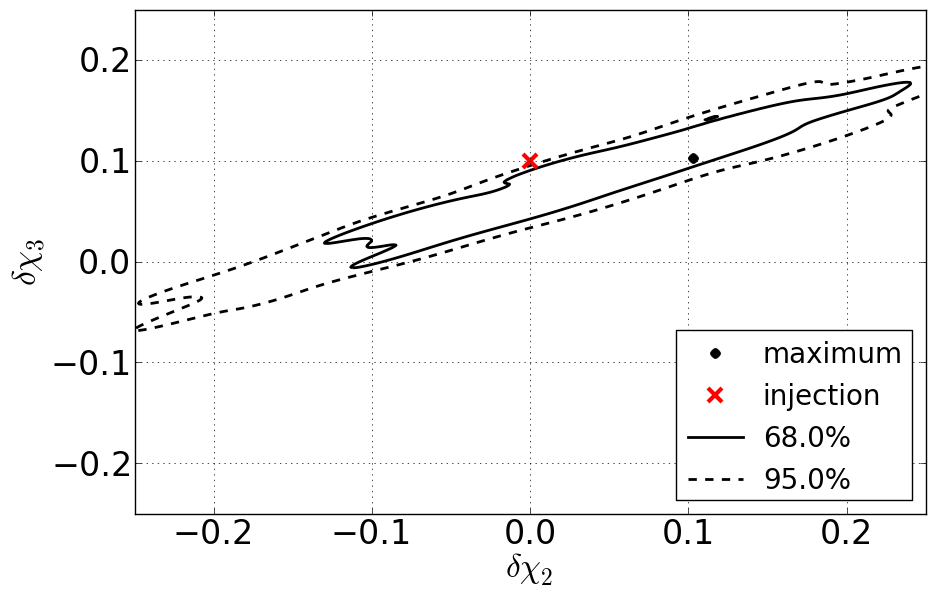}
\caption{The 68$\%$ and 95$\%$ confidence contours of the two-dimensional PDF for $\{\delta\chi_2, \delta\chi_3\}$, still for an injection with $\delta \chi_3 = 0.1$, and with a waveform that has $\{\vec{\theta}, \delta\chi_1, \delta\chi_2, \delta\chi_3\}$ as free parameters. The maximum of the PDF is given by the black dot and the injection values are represented by the red cross.}
\label{fig:2D10pc}
\end{figure}

Finally, we look at the two-dimensional PDF for $\{\delta\chi_2, \delta\chi_3\}$, in the case where the waveform is the one that tests the hypothesis $H_{123}$; Fig.~\ref{fig:2D10pc}. Here too there is little to learn about the underlying nature of the deviation.

\subsubsection{Example: A signal with $\delta\chi_3 = 0.025$}

Let us consider an example with $(\Mc,\eta,D) = (1.14\,M_\odot, 0.242, 216\,\mbox{Mpc})$, $\delta\chi_3 = 0.025$, and a network SNR of 20.6. As expected, the Bayes factors for the component hypotheses against GR are considerably smaller than in the case of $\delta\chi_3 = 0.1$, but GR is still disfavored:
\ba
\ln B^1_{\rm GR} = 11,\,\,\,\,\,\ln B^2_{\rm GR} &=& 12,\,\,\,\,\,\ln B^3_{\rm GR} = 12, \nn\\
\ln B^{12}_{\rm GR} = 10,\,\,\,\,\,\ln B^{13}_{\rm GR} &=& 11,\,\,\,\,\,\ln B^{23}_{\rm GR} = 11, \nn\\
\ln B^{123}_{\rm GR} &=& 11.
\ea
Also as expected, the signal is easily found by all of the model waveforms:
\ba
\ln B^{\rm GR}_{\rm noise} &=& 186,\nn\\
\ln B^1_{\rm noise} = 197,\,\,\,\,\,\ln B^2_{\rm noise} &=& 198,\,\,\,\,\,\ln B^3_{\rm noise} = 198,\nn\\
\ln B^{12}_{\rm noise} = 196,\,\,\,\,\,\ln B^{13}_{\rm noise} &=& 197,\,\,\,\,\,\ln B^{23}_{\rm noise} = 197, \nn\\
\ln B^{123}_{\rm noise} &=& 197. \nn\\
\ea

\begin{figure}[h!]
\centering
\includegraphics[angle=0,width=\columnwidth]{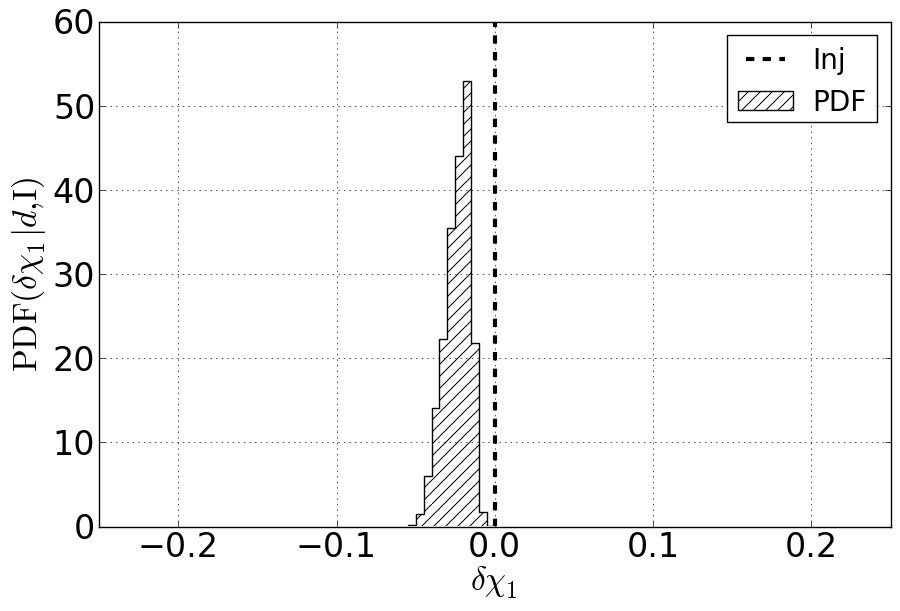}
\includegraphics[angle=0,width=\columnwidth]{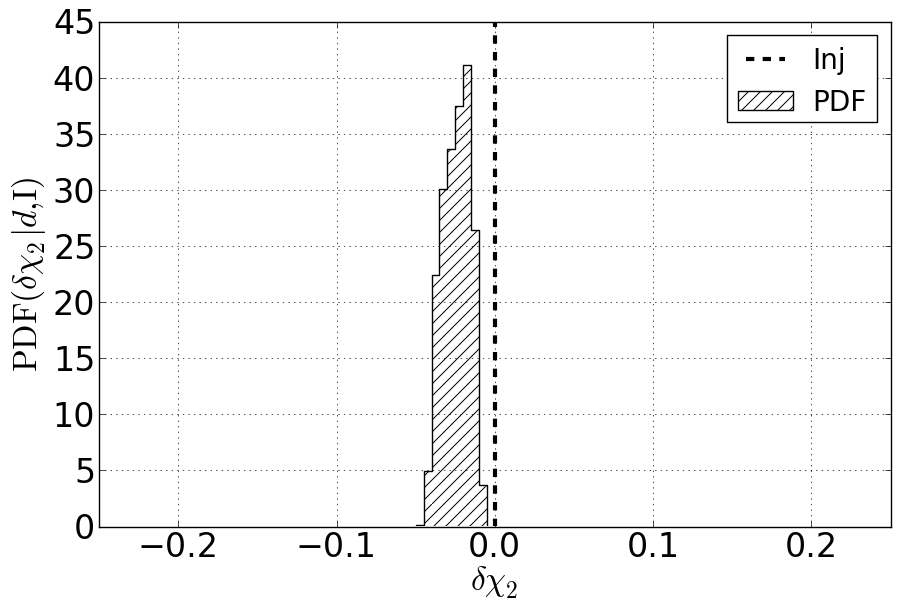}
\includegraphics[angle=0,width=\columnwidth]{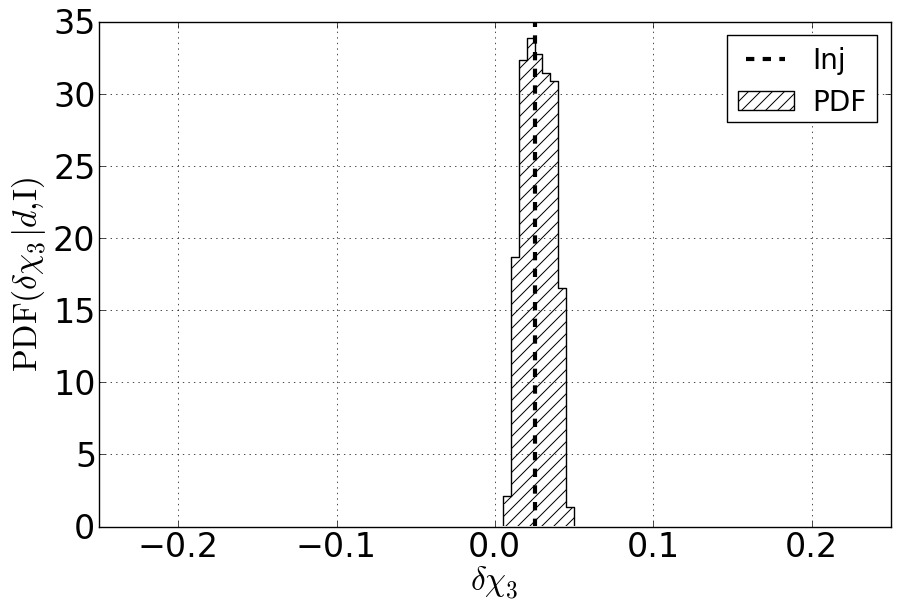}
\caption{The posterior PDFs for $\delta\chi_1$ (top), $\delta\chi_2$ (middle), and $\delta\chi_3$ (bottom) for a single injection with $\delta\chi_3 = 0.025$ and network SNR 20.6, recovered with waveforms where, respectively,
$\{\vec{\theta},\delta\chi_1\}$, $\{\vec{\theta},\delta\chi_2\}$, and  $\{\vec{\theta},\delta\chi_3\}$ are free. Again $\delta\chi_3$ is peaked at close to the correct value, with a median of 0.027 and a standard deviation of 0.0092, but both $\delta\chi_1$ and $\delta\chi_2$ are strongly peaked at incorrect values.}
\label{fig:dpsi3_2p5_dpsi1dpsi2dpsi3}
\end{figure}

As before we look at the posterior PDF of $\delta\chi_3$ for the hypothesis $H_3$, where
only $\{\vec{\theta},\delta\chi_3\}$ are allowed to vary: see the bottom plot in Fig.~\ref{fig:dpsi3_2p5_dpsi1dpsi2dpsi3}. The distribution is
peaked near the correct value and stays away from zero; however, one should not expect the same to happen
for lower-SNR sources.

Let us also look at the PDF for $\delta\chi_1$ when $\{\vec{\theta}, \delta\chi_1\}$ are free parameters, and
of $\delta\chi_2$ when $\{\vec{\theta}, \delta\chi_2\}$ are free; see the top and middle plots of Fig.~\ref{fig:dpsi3_2p5_dpsi1dpsi2dpsi3}. As before,
$\delta\chi_1$ and $\delta\chi_2$ are not peaked at the right values of $\delta\chi_1 = \delta\chi_2 = 0$.

\subsubsection{Example: A signal with non-PN frequency dependence in the phasing}

We now look at a signal with a non-standard contribution to the phase, with a frequency dependence between 1PN and 1.5PN, as in Eq.~(\ref{anomalous}). In the example we use here, $(\Mc,\eta,D) = (1.29\,M_\odot, 0.250, 208\,\mbox{Mpc})$, with a network SNR of 22.4. The Bayes factors for the component hypotheses against GR are:
\ba
\ln B^1_{\rm GR} = 91,\,\,\,\,\,\ln B^2_{\rm GR} &=& 93,\,\,\,\,\,\ln B^3_{\rm GR} = 89, \nn\\
\ln B^{12}_{\rm GR} = 92,\,\,\,\,\,\ln B^{13}_{\rm GR} &=& 91,\,\,\,\,\,\ln B^{23}_{\rm GR} = 92, \nn\\
\ln B^{123}_{\rm GR} &=& 91.
\ea
Thus, also in this case the GR hypothesis is very much disfavored, despite the fact that none of our model waveforms contain the anomalous frequency dependence which is present in the phase of the signal. We can also look at the Bayes factors against noise:
\ba
\ln B^{\rm GR}_{\rm noise} &=& 148,\nn\\
\ln B^1_{\rm noise} = 239,\,\,\,\,\,\ln B^2_{\rm noise} &=& 241,\,\,\,\,\,\ln B^3_{\rm noise} = 237,\nn\\
\ln B^{12}_{\rm noise} = 240,\,\,\,\,\,\ln B^{13}_{\rm noise} &=& 239,\,\,\,\,\,\ln B^{23}_{\rm noise} = 239, \nn\\
\ln B^{123}_{\rm noise} &=& 239. \nn\\
\ea

\begin{figure}[h!]
\centering
\includegraphics[angle=0,width=\columnwidth]{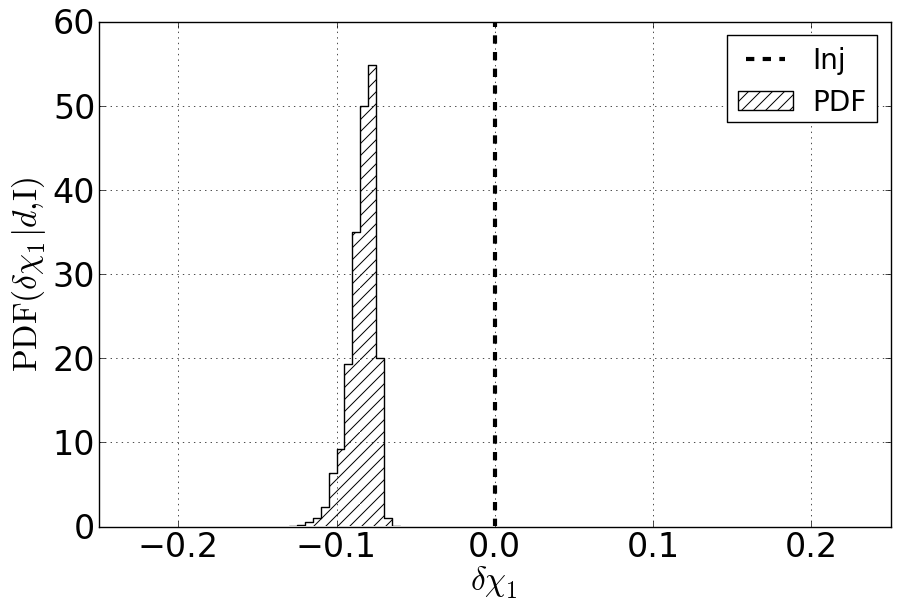}
\includegraphics[angle=0,width=\columnwidth]{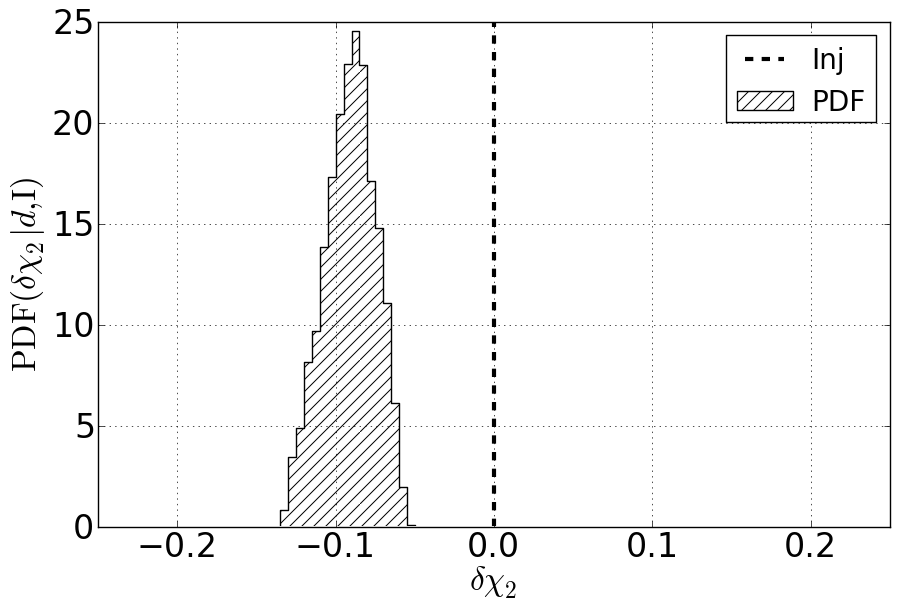}
\includegraphics[angle=0,width=\columnwidth]{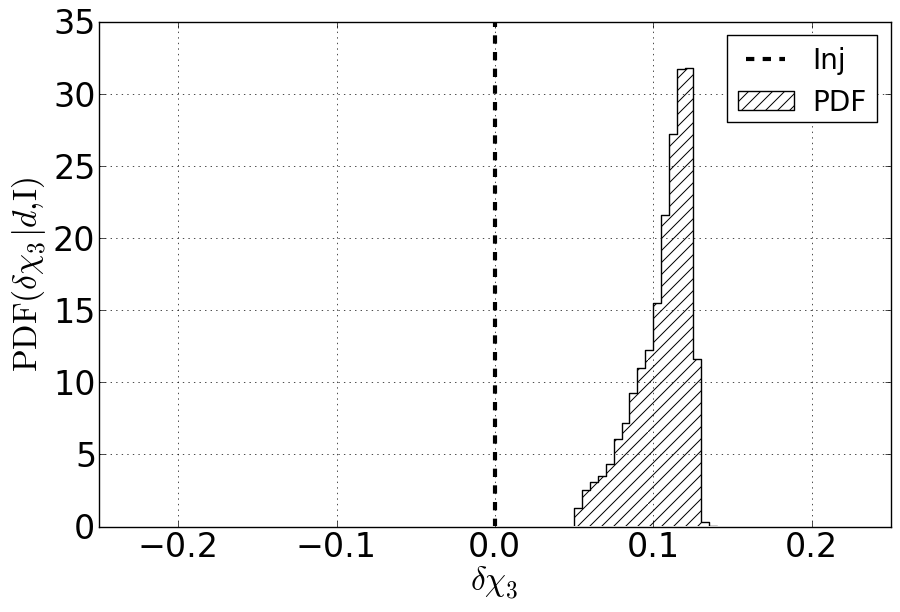}
\caption{The posterior PDFs for $\delta\chi_1$ (top), $\delta\chi_2$ (middle), and $\delta\chi_3$ (bottom) for a single injection with $\delta\chi_A = -2.2$ and network SNR 22.4, recovered with waveforms where, respectively,
$\{\vec{\theta},\delta\chi_1\}$, $\{\vec{\theta},\delta\chi_2\}$, and $\{\vec{\theta},\delta\chi_3\}$ are free. The distribution of $\delta\chi_3$ has its median at 0.11 and a standard deviation of 0.017. Note the remarkable resemblance with Fig.~\ref{fig:dpsi3dpsi3}, where the signal had $\delta\chi_3 = 0.1$. Also for $\delta\chi_1$ and $\delta\chi_2$, the distributions are very similar to the ones for a signal with $\delta\chi_3 = 0.1$.}
\label{fig:dpsiAdpsi1dpsi2dpsi3}
\end{figure}

It is interesting to look at the posterior PDF of $\delta\chi_3$ for the case where $\{\vec{\theta},\delta\chi_3\}$ are allowed to vary (bottom panel of Fig.~\ref{fig:dpsiAdpsi1dpsi2dpsi3}). The distribution looks uncannily like the analogous one for a signal with $\delta\chi_3 = 0.1$; see the bottom panel of Fig.~\ref{fig:dpsi3dpsi3}.
We can also look at the PDF of $\delta\chi_1$ in the case where $\{\vec{\theta},\delta\chi_1\}$ are free parameters, and the PDF of $\delta\chi_2$ when $\{\vec{\theta},\delta\chi_2\}$ are free; see the top and middle panels of Fig.~\ref{fig:dpsiAdpsi1dpsi2dpsi3}. Here too there is an interesting resemblance to the analogous panels in Fig.~\ref{fig:dpsi3dpsi3}, for an injection with $\delta\chi_3 = 0.1$.

In summary,
\begin{itemize}
\item Also here, the Bayes factors against GR clearly disfavor the GR hypothesis, despite the fact that none of our model waveforms has the kind of non-PN contribution to the phase that the signal contains;
\item As before, the Bayes factors against GR are quite close to each other, and one cannot conclude much from them about the nature of the underlying deviation from GR;
\item The Bayes factors against noise indicate that the signal will not be missed;
\item The posteriors are quite similar to the ones where the deviation from GR is purely in the 1.5PN coefficient, with $\delta\chi_3 = 0.1$.
\end{itemize}

\begin{figure}[h!]
\centering
\includegraphics[angle=0,width=\columnwidth]{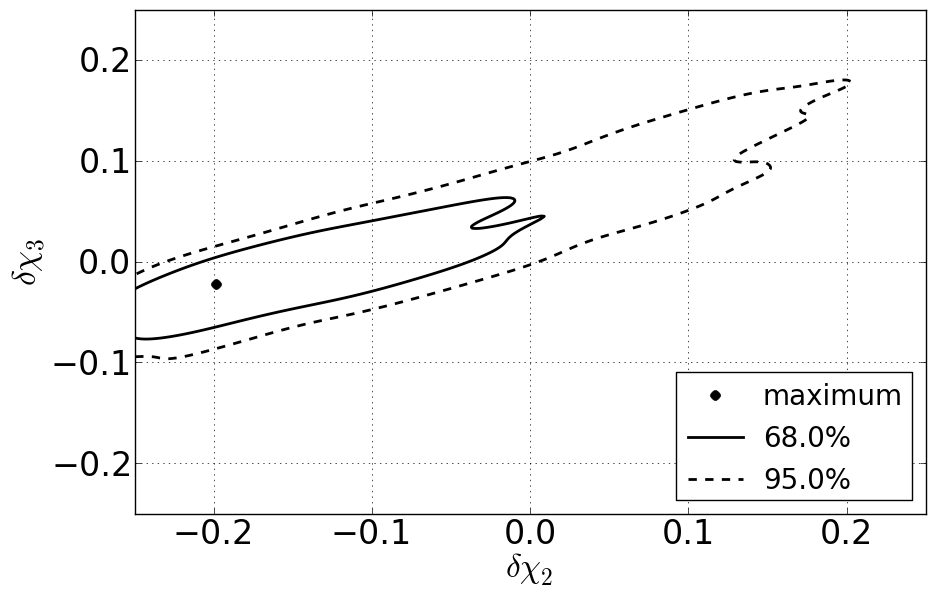}
\caption{The 68$\%$ and 95$\%$ confidence contours of the two-dimensional PDF for $\{\delta\chi_2, \delta\chi_3\}$, for an injection with $\delta \chi_A = -2.2$, and with a waveform that has $\{\vec{\theta}, \delta\chi_1, \delta\chi_2, \delta\chi_3\}$ as free parameters. The maximum of the PDF is given by the black dot. Here the distribution is quite different from the case with $\delta\chi_3 = 0.1$ (Fig.~\ref{fig:2D10pc}).}
\label{fig:2Danomalous}
\end{figure}

Finally, let us look at the two-dimensional PDF for $\{\delta\chi_2, \delta\chi_3\}$ in the case where the waveform is the one that tests $H_{123}$; Fig.~\ref{fig:2Danomalous}. Unlike the one-dimensional PDFs, here there is not much resemblance with the two-dimensional PDF for $\delta\chi_3 = 0.1$ (Fig.~\ref{fig:2D10pc}). Still, nothing much can be learned about the actual nature of the violation.

This shows once again that we will be able to also discern violations of GR of a kind that has no analog in the model waveforms. In the posteriors, these can even `masquerade' as deviations in one of the post-Newtonian coefficients, if, for example, one would only be looking for a `massive graviton' with a waveform model that has $\{\vec{\theta}, \delta\chi_2\}$ as free parameters.

\subsection{A note on parameter estimation and multiple sources}
 
We end this section with some cautionary remarks on the use of parameter estimation in testing GR. As we have seen in subsection \ref{subsec:combining}, model selection allows for combining the information from multiple sources to compute a single odds ratio, but for parameter estimation the situation is quite different.

In the examples presented in this paper, the deviations $\delta\chi_3$ and $\delta\chi_A$ were taken to be the same for all sources. Hence, in principle we could have combined PDFs from multiple sources, simply by multiplying them, to arrive at more accurate measurements. However, we reiterate that in reality, one cannot expect the deviations to be constant in this fashion; rather, they might vary from source to source, with dependence on the masses as well as whatever additional charges may be present in the correct theory of gravity. If it so happens that deviations in individual sources can go both ways, making positive or negative contributions to the overall phase depending on the parameters of the source, then a combined PDF may not show any significant deviation at all. Moreover, since most sources will have SNRs near threshold, it could be that only a few sources will allow for accurate parameter estimation; here we only showed PDFs for single, relatively `loud' sources with SNR $\gtrsim 20$. However, if there is significant dependence of the GR deviation on source parameters (unlike in the heuristic examples shown here), the particular deviation exhibited by the loudest source may not be representative. Hence, in contrast to model selection, parameter estimation alone does not provide a solid foundation to look for deviations from GR.


\section{Conclusions and future directions}

As we showed at the beginning, it is possible for a signal to contain very significant deviations from General Relativity
while still being detectable with a template family of GR waveforms; this is the `fundamental bias' discussed in \cite{yp09,dpvv11}.
A violation of GR can in principle take any form, and the question which then presents itself is how to search for generic
deviations.

We have developed a general method to search for deviations from General Relativity using
signals from compact binary coalescence events. To this end we constructed an odds ratio $O^{\rm modGR}_{\rm GR}$ for
modifications to GR against GR, which is the posterior probability that there is a deviation from
GR, versus GR being correct. This odds ratio can be written as a linear combination of Bayes
factors $B^{i_1 i_2 \ldots i_k}_{\rm GR}$ for hypotheses $H_{i_1 i_2 \ldots i_k}$, in each of which one
or more of the phase parameters $\psi_i$ is assumed to
deviate from the GR value, without actually assuming any specific dependence on
the frequency and/or physical parameters pertinent to a given theory.
Since this includes hypotheses where only a single one of the $\psi_i$ is non-GR, our method will be
particularly well-suited in low-SNR scenarios, which we expect to be in with the upcoming
advanced detector network. Finally, information from multiple sources can easily be combined to 
arrive at an odds ratio $\mathcal{O}^{\rm modGR}_{\rm GR}$ for the `catalog' of all observed events.

The method we developed, applied to phase coefficients in inspiral waveforms, 
essentially addresses the question ``Do \emph{one or more} of the coefficients differ from the values 
predicted by GR?" This is in contrast with previous Bayesian analyses such as \cite{dpvv11,csyp11,gvs11}, 
where effectively, the question being asked was limited to ``Do \emph{all} of the additional free parameters 
introduced differ from their GR values?" In addition to being better adapted to a low-SNR environment,
the test proposed here is far more general.

In order to gauge how large a deviation might be detectable, we first considered signals with a
constant fractional deviation in the 1.5PN coefficient $\psi_3$. This coefficient is of particular interest,
since it incorporates the so-called `tail effects' \cite{bs94,bs95} (as well as spin-orbit coupling \cite{bdiww95}, although we did not consider
spin here), which are not empirically accessible with binary pulsar
observations and can only be studied through direct detection of gravitational waves.
When considering catalogs of only 15 binary neutron star sources, we saw that a deviation in $\psi_3$
at the 10\% level would easily be detectable. In fact, even a deviation at the few percent level can
be discernable. This is confirmed by posterior PDFs for $\psi_3$ in the case where this is the only
parameter that is assumed to deviate from its GR value.

We also considered a deviation in the phase with a frequency dependence that does not match any
of the post-Newtonian terms, and hence is not present in any of the recovery waveforms that we used.
More precisely, we looked at signals whose phase has an additional contribution, with a frequency
dependence in between that of the 1PN and 1.5PN terms (`1.25PN'). The magnitude of the deviation was chosen
such that near $f \sim 150$ Hz, where the detectors are the most sensitive, the change in phase is
roughly the same as the change caused by a 10\% shift in the 1.5PN coefficient. The deviation
was clearly detectable in the log odds ratios, the Bayes factors, and the posterior PDFs. We expect this to
be an instance of a more general fact. Namely, even if there is a deviation in the phase which the
model waveforms technically do not allow for, it will typically be observable, on condition that
near the `bucket' it causes a change in the phase that is on a par with the effect of a shift in
the (low order) PN phase coefficients of more than a few percent.

In order to establish the basic validity and usefulness of our method for testing GR, we considered
constant fractional deviations in $\psi_3$, and a non-PN frequency dependence in the signal. However,
even if there is a deviation in one or more of the PN phase contributions only, these may depend
on $(\Mc,\eta)$ as well as whatever additional charges and coupling constant might be present. It would
be of great interest to study the effects of more general deviations from GR on the
odds ratio $\mathcal{O}^{\rm modGR}_{\rm GR}$, the cumulative Bayes factors $\prod_A {}^{(A)}B^{i_1 i_2 \ldots i_k}_{\rm GR}$
for the component hypotheses $H_{i_1 i_2 \ldots i_k}$ againts $\hyp_{\rm GR}$, and the cumulative Bayes
factors $\prod_A {}^{(A)}B^{i_1 i_2 \ldots i_k}_{\rm noise}$ against noise.
Additionally, in future one should consider deviations in the amplitude as well, \emph{and} use model waveforms which
have such freedom. \emph{A priori} there is no reason not to use
an arbitrary number of free coefficients in both the phase and the amplitude of recovery waveforms, the only limiting factor
 being computational time.

Should no deviation from GR be convincingly found through model selection, one will still be interested in \emph{constraining} the theory. Advanced gravitational wave
detectors will then take us considerably beyond the binary pulsar observations. We recall that even the 1PN phase
coefficient $\psi_2$ is not fully constrained by the latter, since the dissipative dynamics is only probed to leading PN
order. As we have seen from PDFs, with a single compact binary coalescence event in Advanced LIGO/Virgo at SNR $\sim 20$,
 the coefficient $\psi_2$ can be constrained to better than 2\%, and the same is true of the 1.5PN coefficient $\psi_3$.
Mainly for computational reasons, we did not study constraints on the 2PN and higher-order terms, but it would clearly
be of interest to see how well one can pin down the corresponding coefficients. We also draw attention to the
results for the 0.5PN contribution, which in General Relativity is identically zero. For simplicity
we restricted the mass range of our simulated sources to that of binary neutron stars. Given the very encouraging results,
in future work the BHNS and BBH ranges should also be investigated.

The preliminary results presented here motivate the construction of a full data analysis pipeline for testing
or constraining General Relativity. To have a real chance of finding a deviation, much more sophisticated GR waveforms
will have to be used, with inclusion of merger and ringdown, higher harmonics both in the inspiral and ringdown
parts, dynamical spins, and residual eccentricity. The development of such waveforms, with input from numerical
simulations, is currently a subject of intense investigations
\cite{ajith07,ajith11,ajith10,bd99,bd00,djs00,Damour:2001,bcd06,buonanno11,rs10a,rs10b}. We note that the method we presented is not tied to any particular waveform approximant. Moreover, the deformations need not be in the phase. Indeed, in the future one would presumably want to use time domain waveforms, for which it may be more convenient (and physically more appropriate) to introduce parameterized deformations directly in coefficients appearing in, \emph{e.g.}, a Hamiltonian used to evolve the inspiral part of the waveform. Irrespective of the parameterization, one would still be able to associate with it an exhaustive set of logically disjoint hypotheses $H_{i_1 i_2 \ldots i_k}$.

Once sufficiently accurate
waveforms are available, a test of GR on Advanced LIGO/Virgo could go as follows:
\begin{itemize}
\item Starting from the best available GR waveforms, introduce parameterized deformations,
leading to disjoint hypotheses like our $H_{i_1 i_2 \ldots i_k}$, which together form $\hyp_{\rm modGR}$;
\item Use many injections of GR waveforms in real or realistic data, arranged into simulated `catalogs', to investigate
the distributions of the cumulative odds ratio
$\mathcal{O}^{\rm modGR}_{\rm GR}$ as well as of the cumulative Bayes factors when GR is correct. Use these to set \emph{thresholds}
for the \emph{measured} odds ratio and Bayes factors to overcome;
\item Apply our method to the catalog of sources actually found by the detectors. If the measured cumulative odds ratio $\mathcal{O}^{\rm modGR}_{\rm GR}$
is below threshold, then there is no real reason to believe that a deviation from GR is present. The posterior PDFs for the free
phase and amplitude coefficients in the model waveforms, taken from the highest-SNR sources, will provide (potentially very strong)
constraints on these parameters;
\item If the measured odds ratio is above threshold, then a violation of GR is likely. As we have seen, Bayes factors
and PDFs can be misleading in trying to find out what the precise nature of the deviation may be. However, one may be able to follow up
on the violation by again using our method, this time with waveforms with more complicated deformations and a larger number of
free parameters, inspired by particular alternative theories of gravity, similar to what is done in ppE \cite{yp09}.
\end{itemize}

Thus, although much work remains to be done, we have the basics of a very general method for testing General Relativity using
compact binary coalescence events to be detected by the upcoming Advanced LIGO and Advanced Virgo observatories.

\section*{Acknowledgements}

TGFL, WDP, SV, CVDB and MA are supported by the research programme of the Foundation for Fundamental Research on Matter (FOM), which is partially supported by the Netherlands Organisation for Scientific Research (NWO). JV's research was funded by the Science and Technology Facilities Council (STFC), UK, grant ST/J000345/1. KG, TS and AV are supported by Science and Technology Facilities Council (STFC), UK, grant ST/H002006/1. The work of RS is supported by the EGO Consortium through the VESF fellowship EGO-DIR-41-2010.

It is a pleasure to thank N.~Cornish, B.R.~Iyer, B.S.~Sathyaprakash, and N.~Yunes for their valuable comments and suggestions. We also thank the anonymous referee, whose comments helped to greatly improve the presentation of the paper. The authors would like to acknowledge the LIGO Data Grid clusters, without which the simulations could not have been performed. Specifically, these include the computing resources supported by National Science Foundation awards PHY-0923409 and PHY-0600953 to UW-Milwaukee. Also, we thank the Albert Einstein Institute in Hannover, supported by the Max-Planck-Gesellschaft, for use of the Atlas high-performance computing cluster.



%
%


\begin{thebibliography}{99}

\bibitem{MTW} C.W.~Misner, K.S.~Thorne, and J.A.~Wheeler, \emph{Gravitation}, W.H.~Freeman and Company, New York, 1973

\bibitem{Will:2006}
 {C.~M.~Will}, Liv.~Rev.~Rel.~{\bf 9}, 3 (2006)
  \url{http://www.livingreviews.org/lrr-2006-3}

\bibitem{SathyaprakashSchutz:2009}
  B.~S.~Sathyaprakash and B.~F.~Schutz,
  Living Rev.\ Rel.\  {\bf 12}, 2 (2009)
  [arXiv:0903.0338 [gr-qc]].

\bibitem{ht75} R.A.~Hulse and J.H.~Taylor, Astrophys.~J.~{\bf 195}, L51 (1975)

\bibitem{b03} M.~Burgay \emph{et al.}, Nature {\bf 426}, 531 (2003); astro-ph/0312071

\bibitem{k06} M.~Kramer \emph{et al.}, Science {\bf 314}, 97 (2006); astro-ph/0609417

\bibitem{LIGO} B.~Abbott \emph{et al.}, Rep.~Prog.~Phys. {\bf 72}, 076901 (2009)

\bibitem{virgo}
F.~Acernese \emph{et al.}, Class.\ Quant.\ Grav. {\bf 21},  385 (2004)

\bibitem{Virgo} F.~Acernese \emph{et al.}, Class.Quantum Grav. {\bf 25}, 184001 (2008); T.~Accadia \emph{et al.}, Class.~Quantum Grav.~{\bf 28}, 114002 (2011)

\bibitem{GEO600} H.~Grote (for the LIGO Scientific Collaboration), Class.~Quantum Grav.~{\bf 25}, 114043 (2008); H Grote and the LIGO Scientific Collaboration, Class~ Quantum Grav.~{\bf 27}, 084003 (2010)

\bibitem{aLIGO} G.M.~Harry (for the LIGO Scientific Collaboration), Class.~Quantum Grav.~{\bf 27} 084006 (2010)

\bibitem{advligo}
\url{http://www.ligo.caltech.edu/advLIGO/}

\bibitem{aVirgo} Advanced Virgo Baseline Design, The Virgo Collaboration, note VIR–027A–09 May 16, 2009
https://tds.ego-gw.it/itf/tds/file.php?callFile=VIR-0027A-09.pdf

\bibitem{advvirgo}
\url{http://wwwcascina.virgo.infn.it/advirgo/}

\bibitem{ratespaper}
  J.~Abadie {\it et al.}  [LIGO Scientific Collaboration and Virgo
                  Collaboration],
  Class.\ Quant.\ Grav.\  {\bf 27}, 173001 (2010); arXiv:1003.2480

\bibitem{LCGT} K.~Kuroda and the LCGT Collaboration, Class.~Quantum Grav.~{\bf 27}, 084004 (2010)


\bibitem{IndiGO} B.S.~Sathyaprakash for the LIGO Scientific Collaboration, \emph{Scientific Benefits of LIGO-India}, LSC internal report G1100991 (2011)

\bibitem{PN} L.~Blanchet, Liv.~Rev.~Rel.~{\bf 5}, 3 (2002); \url{http://relativity.livingreviews.org/Articles/lrr-2006-4/}

\bibitem{Maggiore} M.~Maggiore, \emph{Gravitational Waves. Volume 1: Theory and Experiments}, Oxford University Press, Oxford, 2008

\bibitem{bs94} L.~Blanchet and B.S.~Sathyaprakash, Class.~Quantum Grav.~{\bf 11}, 2807 (1994)

\bibitem{bs95} L.~Blanchet and B.S.~Sathyaprakash, Phys.~Rev.~Lett.~{\bf 74}, 1067 (1995)

\bibitem{bdiww95} L.~Blanchet, T.~Damour, B.~Iyer, C.M.~Will, A.G.~Wiseman, Phys.~Rev.~Lett.~{\bf 74}, 3515-3518 (1995)

\bibitem{mais10} C.K.~Mishra, K.G.~Arun, B.R.~Iyer, and B.S.~Sathyaprakash, Phys.~Rev.~D {\bf 82}, 064010 (2010); arXiv:1005.0304

\bibitem{w94} C.M.~Will, Phys.~Rev.~D {\bf 50}, 6058-6067 (1994); gr-qc/9406022

\bibitem{de98} T.~Damour and G.~Esposito-Far\`ese , Phys.~Rev.~D {\bf 58}, 042001 (1998)

\bibitem{sw01} P.D.~Scharre and C.M.~Will, Phys.~Rev.~D {\bf 65}, 042002 (2002)

\bibitem{wy04} C.M.~Will and N.~Yunes, Class.~Quant.~Grav.~{\bf 21}, 4367 (2004); gr-qc/0403100

\bibitem{bbw04} E.~Berti, A.~Buonanno, and C.M.~Will, Phys.~Rev.~D {\bf 71}, 084025 (2005); gr-qc/0411129

\bibitem{bbw05} E.~Berti, A.~Buonanno, and C.M.~Will, Class.~Quantum Grav.~{\bf 22}, S943-S954 (2005); gr-qc/0504117

\bibitem{yps10} N.~Yunes, F.~Pretorius, and D.~Spergel, Phys.~Rev.~D {\bf 81}, 064018 (2010); arXiv:0912.2724

\bibitem{w98} C.M.~Will, Phys.~Rev.~D {\bf 57}, 2061-2068 (1998); gr-qc/9709011

\bibitem{aw09} K.G.~Arun and C.M.~Will, Class.~Quantum Grav.~{\bf 26}, 155002 (2009); arXiv:0904.1190

\bibitem{sw09} A.~Stavridis and C.M.~Will, Phys.~Rev.~D {\bf 80}, 044002 (2009)

\bibitem{ka10} D.~Keppel and P.~Ajith, Phys.~Rev.~D~{\bf 82},122001 (2010); arXiv:1004.0284

\bibitem{dpvv11} W.~Del Pozzo, J.~Veitch, and A.~Vecchio, Phys.~Rev.~D {\bf 83}, 082002 (2011); arXiv:1101.1391

\bibitem{BertiGairSesana:2011}
  E.~Berti, J.~Gair and A.~Sesana; arXiv:1107.3528 (2011)

\bibitem{Ryan:1997}
  F.~D.~Ryan,
  Phys.\ Rev.\  D {\bf 56}, 1845 (1997)

\bibitem{bcw06} E.~Berti, V.~Cardoso, and C.M.~Will, Phys.~Rev.~D {\bf 73}, 064030 (2006); gr-qc/0512160

\bibitem{Hughes:2006}
  S.~A.~Hughes,
  AIP Conf.\ Proc.\  {\bf 873}, 233 (2006); gr-qc/0608140

\bibitem{bc07} L.~Barack and C.~Cutler, Phys.~Rev.~D {\bf 75}, 042003 (2007); gr-qc/0612029

\bibitem{glm08} J.R.~Gair, C.~Li, I.~Mandel, Phys.~Rev.~D {\bf 77}, 024035 (2008); arXiv:0708.0628

\bibitem{khhs11} I.~Kamaretsos, M.~Hannam, S.~Husa, B.S.~Sathyaprakash; arXiv:1107.0854 (2011)

\bibitem{vdbs06} C.~Van Den Broeck and A.S.~Sengupta, Class.~Quantum Grav.~{\bf 24}, 1089-1113 (2007); gr-qc/0610126

\bibitem{afy08} S.~Alexander, L.S.~Finn, and N.~Yunes, Phys.~Rev.~D {\bf 78}, 066005 (2008); arXiv:0712.2542

\bibitem{yf09} N.~Yunes and L.S.~Finn, J.~Phys.~Conf.~Ser.~{\bf 154}, 012041 (2009); arXiv:0811.0181

\bibitem{sy09} C.~Sopuerta and N.~Yunes, Phys.~Rev.~D {\bf 80}, 064006 (2009)

\bibitem{yooa10} N.~Yunes, R.~O'Shaughnessy, B.J.~Owen, S.~Alexander,  Phys.~Rev.~D {\bf 82}, 064017 (2010); arXiv:1005.3310

\bibitem{aiqs06a} K.G.~Arun, B.R.~Iyer, M.S.S.~Qusailah, and B.S.~Sathyaprakash, Class.~Quantum Grav.~{\bf 23}, L37-L43 (2006); 0604018

\bibitem{aiqs06b} K.G.~Arun, B.R.~Iyer, M.S.S.~Qusailah, and B.S.~Sathyaprakash, Phys.~Rev.~D {\bf 74}, 024006 (2006); gr-qc/0604067

\bibitem{yp09} N.~Yunes and F.~Pretorius, Phys.~Rev.~D {\bf 80}, 122003 (2009); arXiv:0909.3328

\bibitem{yh10} N.~Yunes and S.A.~Hughes, Phys.~Rev.~D {\bf 82}, 082002 (2010); arXiv:1007.1995

\bibitem{csyp11} N.~Cornish, L.~Sampson, N.~Yunes, and F.~Pretorius, arXiv:1105.2088 (2011)

\bibitem{gvs11} S.~Gossan, J.~Veitch, and B.S.~Sathyaprakash, arXiv:1111.5819 (2011)

\bibitem{bfis08} L.~Blanchet, G.~Faye, B.R.~Iyer, and S.~Sinha, Class.~Quantum~Grav.~{\bf 25}, 165003 (2008); arXiv:0802.1249

\bibitem{buonanno11} Y.~Pan \emph{et al.}, arXiv:1107.2904 (2011)

\bibitem{vdbs07} C.~Van Den Broeck and A.S.~Sengupta, Class.~Quantum Grav.~{\bf 24}, 155-176 (2007); gr-qc/0607092

\bibitem{vz10} S.~Vitale and M.~Zanolin, Phys.~Rev.~D {\bf 82}, 124065 (2010); arXiv:1004.4537

\bibitem{thorne87} K.S.~Thorne, in \emph{300 Years of Gravitation}, eds. S.W. Hawking and W. Israel, Cambridge
University Press, Cambridge, England, 1987

\bibitem{sd91} B.S.~Sathyaprakash and S.V.~Dhurandhar, Phys.~Rev. D {\bf 44}, 3819-3934 (1991)

\bibitem{LAL} \url{https://www.lsc-group.phys.uwm.edu/daswg/projects/lalsuite.html}

\bibitem{biops09} A.~Buonanno, B.~Iyer, E.~Ochsner, Y.~Pan, and B.S.~Sathyaprakash, Phys.~Rev.~D {\bf 80}, 084043 (2009); arXiv:0907.0700

\bibitem{AdvLIGOnoise} D.~Shoemaker (LSC, 2009), \url{https://dcc.ligo.org/cgi-bin/DocDB/ShowDocument?docid=2974}

\bibitem{apostolatos95} T.~A.~Apostolatos,~ Phys.~Rev.~D {\bf 52}, 605 (1995)

\bibitem{Skilling:AIP}
J.~Skilling, in \emph{AIP Conference Proceedings:
24th International Workshop on Bayesian Inference and Maximum Entropy Methods in Science and Engineering}, Volume 735 pp.\,395-405 (2004)

\bibitem{VeitchVecchio:2008a}
  J.~Veitch and A.~Vecchio,
  Phys.\ Rev.\  D {\bf 78}, 022001 (2008); arXiv:0801.4313

\bibitem{VeitchVecchio:2008b}
  J.~Veitch and A.~Vecchio,
  Class.\ Quant.\ Grav.\  {\bf 25}, 184010 (2008); arXiv:0807.4483

\bibitem{vv09} J.~Veitch and A.~Vecchio, Phys.~Rev.~D {\bf 81}, 062003 (2010); arXiv:0911.3820

\bibitem{AylottEtAl:2009}
  B.~Aylott {\it et al.},
  Class.\ Quant.\ Grav.\  {\bf 26}, 165008 (2009); arXiv:0901.4399

\bibitem{AVV:2009}
  B.~Aylott , J.~Veitch and A.~Vecchio,
  Class.\ Quant.\ Grav.\  {\bf 26}, 114011 (2009)

\bibitem{vdvz} H.~van~Dam and M.J.G.~Veltman, Nucl. Phys. B {\bf 22}, 397 (1970); V.I.~Zhakarov JETP Lett. {\bf 12} 312 (1970).

\bibitem{deRham:2010kj}
  C.~de Rham, G.~Gabadadze and A.~J.~Tolley,
  Phys.\ Rev.\ Lett.\  {\bf 106} (2011) 231101
  [arXiv:1011.1232 [hep-th]].

\bibitem{ajith07} P.~Ajith \emph{et al.}, Class.~Quantum Grav.~{\bf 24}, S689-S700 (2007)

\bibitem{ajith11} P.~Ajith \emph{et al.}, Phys.~Rev.~Lett.~{\bf 106}, 241101 (2011); arXiv:0909.2867

\bibitem{ajith10} L.~Santamaria \emph{et al.}, Phys.~Rev.~D {\bf 82}, 064016 (2010); arXiv:1005.3306

\bibitem{bd99} A.~Buonanno and T.~Damour, Phys.~Rev.~D {\bf 59}, 084006 (1999)

\bibitem{bd00} A.~Buonanno and T.~Damour, Phys.~Rev.~D {\bf 62}, 064015 (2000)

\bibitem{djs00} T.~Damour, P.~Jaranowski, and G.~Sch\"{a}fer, Phys.~Rev.~D {\bf 62}, 084011 (2000)

\bibitem{Damour:2001} T.~Damour, Phys.~Rev.~D {\bf 64}, 124013 (2001)

\bibitem{bcd06} A.~Buonanno, Y.~Chen, and T.~Damour, Phys.~Rev.~D {\bf 74}, 104005 (2006)

\bibitem{rs10a} R.~Sturani, S.~Fischetti, L.~Cadonati, G.M.~Guidi, J.~Healy, and D.~Shoemaker,
J.~Phys.~Conf.~Ser.~{\bf 243}, 012007 (2010)

\bibitem{rs10b} R.~Sturani, S.~Fischetti, L.~Cadonati, G.M.~Guidi, J.~Healy, and D.~Shoemaker (2011); arXiv:1012.5172

\end{thebibliography}
\end{document}